%% file: HIN-12-016_temp.tex
\pdfoutput=1

\documentclass[11pt,twoside,a4paper,cmspaper,final,collab]{cms-tdr}
\def\svnVersion{246097}\def\cmsCernNo{2013-096}\def\cmsCernDate{\today}\def\cmsMessage{Published in the European Physical Journal C as \href{http://dx.doi.org/10.1140/epjc/s10052-014-2847-x}{\doi{10.1140/epjc/s10052-014-2847-x.}}}

\begin{document}\cmsNoteHeader{HIN-12-016}

\hyphenation{had-ron-i-za-tion}
\hyphenation{cal-or-i-me-ter}
\hyphenation{de-vices}

\RCS$Revision: 232434 $
\RCS$HeadURL: svn+ssh://svn.cern.ch/reps/tdr2/papers/HIN-12-016/trunk/HIN-12-016.tex $
\RCS$Id: HIN-12-016.tex 232434 2014-03-18 22:00:45Z sikler $
\newlength\cmsFigWidth
\ifthenelse{\boolean{cms@external}}{\setlength\cmsFigWidth{0.32\textwidth}}{\setlength\cmsFigWidth{0.49\textwidth}}
\ifthenelse{\boolean{cms@external}}{\providecommand{\cmsLeft}{top\xspace}}{\providecommand{\cmsLeft}{left\xspace}}
\ifthenelse{\boolean{cms@external}}{\providecommand{\cmsRight}{bottom\xspace}}{\providecommand{\cmsRight}{right\xspace}}
\ifthenelse{\boolean{cms@external}}{\providecommand{\cmsLLeft}{Top\xspace}}{\providecommand{\cmsLLeft}{Left\xspace}}
\ifthenelse{\boolean{cms@external}}{\providecommand{\cmsRRight}{Bottom\xspace}}{\providecommand{\cmsRRight}{Right\xspace}}
\ifthenelse{\boolean{cms@external}}{\providecommand{\breakhere}{\linebreak[4]}}{\providecommand{\breakhere}{\relax}}

\newcommand{\mpe}{\ensuremath{\ln\varepsilon}\xspace}
\newcommand{\nh}{\ensuremath{{n_\text{hits}}}\xspace}
\newcommand{\mt}{\ensuremath{m_{\mathrm{T}}}\xspace}

\renewcommand{\HIJING}{\textsc{Hijing}\xspace}
\newcommand{\AMPT}{\textsc{Ampt}\xspace}
\newcommand{\EPOSLHC}{\textsc{Epos Lhc}\xspace}
\newcommand{\EPOS}{\textsc{Epos}\xspace}

\newlength\sptw\setlength\sptw{0.07in}

\cmsNoteHeader{HIN-12-016}

\title{Study of the production of charged pions, kaons, and protons in pPb collisions at $\sqrt{s_{NN}} =\; $5.02\TeV}

\date{\today}

\abstract{
Spectra of identified charged hadrons are measured in pPb collisions with the
CMS detector at the LHC at $\sqrt{s_{NN}} =5.02\TeV$. Charged pions, kaons, and
protons in the transverse-momentum range $\pt\approx0.1$--1.7\GeVc and
laboratory rapidity $\abs{y} < 1$ are identified via their energy loss in the
silicon tracker.  The average \pt increases with particle mass and the charged
multiplicity of the event. The increase of the average \pt with charged
multiplicity is greater for heavier hadrons.  Comparisons to Monte Carlo event
generators reveal that \EPOSLHC, which incorporates additional hydrodynamic
evolution of the created system, is able to reproduce most of the data
features, unlike \HIJING and \AMPT. The \pt spectra and integrated yields are
also compared to those measured in pp and PbPb collisions at various energies.
The average transverse momentum and particle ratio measurements indicate that
particle production at LHC energies is strongly correlated with event particle
multiplicity.
}

\hypersetup{%
pdfauthor={CMS Collaboration},%
pdftitle={Study of the production of charged pions, kaons, and
protons in pPb collisions at sqrt(sNN) = 5.02 TeV},%
pdfsubject={CMS},%
pdfkeywords={CMS, physics, energy loss, hadron spectra}}

\maketitle

\section{Introduction}

\label{sec:introduction}

The study of hadron production has a long history in high-energy particle and
nuclear physics, as well as in cosmic-ray physics.
The absolute yields and the transverse momentum ($\pt$) spectra of identified
hadrons in high-energy hadron-hadron collisions are among the most basic
physical observables. They can be used to test the predictions for
non-perturbative quantum chromodynamics (QCD) processes like hadronization and
soft-parton interactions, and the validity of their implementation in Monte
Carlo (MC) event generators.
Spectra of identified particles in proton-nucleus collisions also constitute an
important reference for studies of high-energy heavy-ion collisions, where
final-state effects are known to modify the spectral shape and yields of
different hadron
species~\cite{Schnedermann:1993ws,Huovinen:2006jp,Adler:2003cb,Arsene:2005mr,Back:2006tt,Abelev:2008ab,Adare:2013esx}.

The present analysis focuses on the measurement of the $\pt$ spectra of charged
hadrons, identified mostly via their energy deposits in silicon detectors, in
pPb collisions at $\sqrt{s_{NN}} =$ 5.02\TeV.
The analysis procedures are similar to those previously used in the measurement
of pion, kaon, and proton production in pp collisions at several center-of-mass
energies~\cite{identifiedSpectra}.
Results on \Pgp, \PK, and \Pp\ production in
pPb collisions have been also reported by the ALICE Collaboration~\cite{Abelev:2013haa}.

A detailed description of the CMS (Compact Muon Solenoid) detector can be found
in Ref.~\cite{:2008zzk}.
The CMS experiment uses a right-handed coordinate system, with the origin at
the nominal interaction point (IP) and the $z$ axis along the
counterclockwise-beam direction.
The pseudorapidity $\eta$ and rapidity $y$ of a particle (in the laboratory
frame) with energy $E$, momentum $p$, and momentum along the $z$ axis $p_z$ are
defined as $\eta = -\ln[\tan(\theta/2)]$, where $\theta$ is the polar angle
with respect to the $z$ axis and $y = \frac{1}{2}\ln[(E+p_z)/(E-p_z)]$,
respectively.
The central feature of the CMS apparatus is a superconducting solenoid of
6\unit{m} internal diameter. Within the 3.8 T field volume are the silicon
pixel and strip tracker, the crystal electromagnetic calorimeter, and the
brass/scintillator hadron calorimeter.
The tracker measures charged particles within the pseudorapidity range
$\abs{\eta} < 2.4$. It has 1440 silicon pixel and 15\,148 silicon strip
detector modules, ordered in 13 tracking layers in the $y$ region studied here.
In addition to the barrel and endcap detectors, CMS has extensive forward
calorimetry.
Steel/quartz-fiber forward calorimeters (HF) cover $3 < \abs{\eta} < 5$.
Beam Pick-up Timing for the eXperiments (BPTX) devices were used to trigger the
detector readout. They are located around the beam pipe at a distance of
175\unit{m} from the IP on either side, and are designed to provide precise
information on the Large Hadron Collider (LHC) bunch structure and timing of
the incoming beams.

The reconstruction of charged particles in CMS is bounded by the acceptance of
the tracker ($\abs{\eta} < $ 2.4) and by the decreasing tracking efficiency at
low momentum (greater than about 60\% for $p > 0.05$, 0.10, 0.20, and 0.40\GeVc
for \Pe, \Pgp, \PK, and \Pp, respectively). Particle identification
capabilities using specific ionization are restricted to $p < 0.15\GeVc$ for
electrons, $p < 1.20\GeVc$ for pions, $p < 1.05\GeVc$ for kaons, and $p <
1.70\GeVc$ for protons.
Pions are identified up to a higher momentum than kaons because of their high
relative abundance.
In view of the $(y,\pt)$ regions where pions, kaons, and protons can all be
identified ($p = \pt \cosh y$), the band $-1 < y < 1$ (in the laboratory frame)
was chosen for this measurement, since it is a good compromise between the \pt
range and $y$ coverage.

In this paper, comparisons are made to predictions from three MC event
generators.
The \HIJING~\cite{Deng:2010mv} event generator is based on a two-component
model for hadron production in high-energy nucleon and nuclear collisions.
Hard parton scatterings are assumed to be described by perturbative QCD and
soft interactions are approximated by string excitations with an effective
cross section. In version 2.1~\cite{Xu:2012au}, in addition to modification of
initial parton distributions, multiple scatterings inside a nucleus lead to
transverse momentum broadening of both initial and final-state partons. This is
responsible for the enhancement of intermediate-$\pt$ (2--6\GeVc) hadron
spectra in proton-nucleus collisions, with respect to the properly scaled
spectra of proton-proton collisions (Cronin effect).
The \AMPT \cite{Lin:2011zzg} event generator is a multi-phase transport model.
It starts from the same initial conditions as \HIJING, contains a partonic
transport phase, the description of the bulk hadronization, and finally a
hadronic rescattering phase.
These processes lead to hydrodynamic-like effects in simulated nucleus-nucleus
collisions, but not necessarily in proton-nucleus collisions.
The latest available version (1.26/2.26) is used.
The \EPOS~\cite{Werner:2005jf} event generator uses a quantum mechanical
multiple scattering approach based on partons and strings, where cross sections
and particle production are calculated consistently, taking into account energy
conservation in both cases. Nuclear effects related to transverse momentum
broadening, parton saturation, and screening have been introduced. The model
can be used both for extensive air shower simulations and accelerator physics.
\EPOSLHC~\cite{Pierog:2013ria} is an improvement of version 1.99 (v3400) and
contains a three-dimensional viscous event-by-event hydrodynamic treatment.
This is a major difference with respect to the \HIJING and \AMPT models for
proton-nucleus collisions.

\section{Data analysis}

\label{sec:dataAnalysis}

The data were taken in September 2012 during a 4-hour-long pPb run with very
low probability of multiple interactions (0.15\% ``pileup''). A total of 2.0
million collisions were collected, corresponding to an integrated luminosity of
approximately $1\mubinv$.
The dominant uncertainty for the reported measurements is systematic in nature.
The beam energies were 4\TeV for protons and 1.58\TeV per nucleon for lead
nuclei, resulting in a center-of-mass energy per nucleon pair of $\sqrt{s_{NN}}
=$ 5.02\TeV.
Due to the asymmetric beam energies the nucleon-nucleon center-of-mass in the
pPb collisions was not at rest with respect to the laboratory frame but was
moving with a velocity $\beta = -0.434$ or rapidity $-0.465$. Since the
higher-energy proton beam traveled in the clockwise direction, i.e. at $\theta
= \pi$, the rapidity of a particle emitted at $y_\text{cm}$ in the
nucleon-nucleon center-of-mass frame is detected in the laboratory frame with a
shift, $y - y_\text{cm} = -0.465$, i.e. a particle with $y=0$ moves with
rapidity 0.465 in the Pb-beam direction in the center-of-mass system.
The particle yields reported in this paper have been measured for laboratory
rapidity $\abs{y} < 1$ to match the experimentally accessible region.

The event selection consisted of the following requirements:

\begin{itemize}

 \item at the trigger level, the coincidence of signals from both BPTX devices,
indicating the presence of both proton and lead bunches crossing the
interaction point; in addition, at least one track with $\pt > 0.4\GeVc$ in the
pixel tracker;

 \item offline, the presence of at least one tower with energy above 3\GeV in
each of the HF calorimeters; at least one reconstructed interaction vertex;
beam-halo and beam-induced background events, which usually produce an
anomalously large number of pixel hits~\cite{Khachatryan:2010xs}, are
suppressed.

\end{itemize}

The efficiencies for event selection, tracking, and vertexing were evaluated
using simulated event samples produced with the \HIJING 2.1 MC event generator,
where the CMS detector response simulation was based on
\GEANTfour \cite{Agostinelli2003250}. Simulated events were reconstructed in
the same way as collision data events.
The final results were corrected to a particle level selection applied to the
direct MC output, which is very similar to the data selection described above:
at least one particle (proper lifetime $\tau > 10^{-18}\unit{s}$) with $E >
3\GeV$ in the range $-5 < \eta < -3$ and at least one in the range $3 < \eta
<5$; this selection is referred to in the following as the ``double-sided''
(DS) selection. These requirements are expected to suppress single-diffractive
collisions in both the data and MC samples.
From the MC event generators studied in this paper, the DS selection efficiency
for inelastic, hadronic collisions is found to be 94--97\%.

\begin{figure}[htbp]

 \begin{center}
  \includegraphics[width=0.49\textwidth]{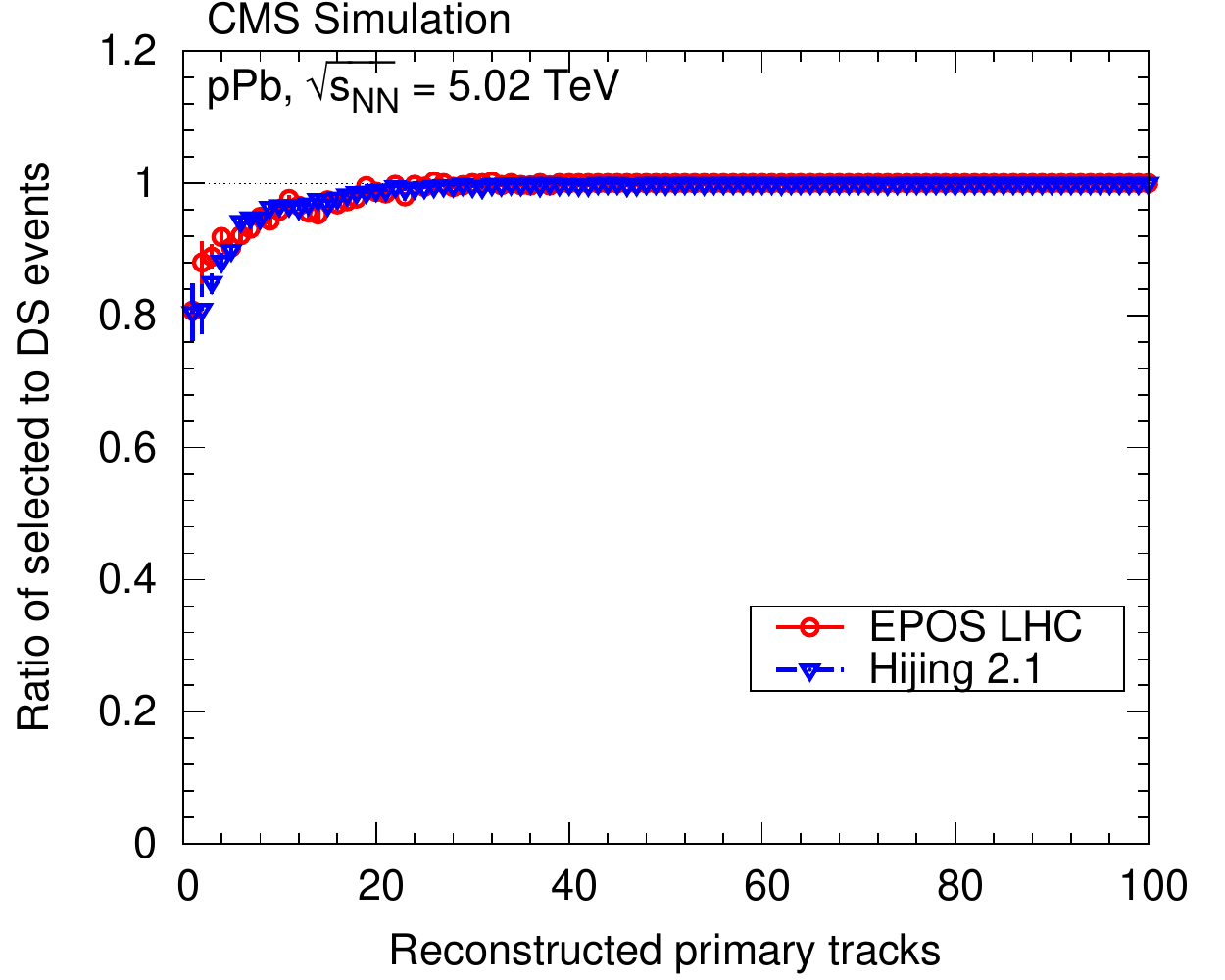}
  \includegraphics[width=0.49\textwidth]{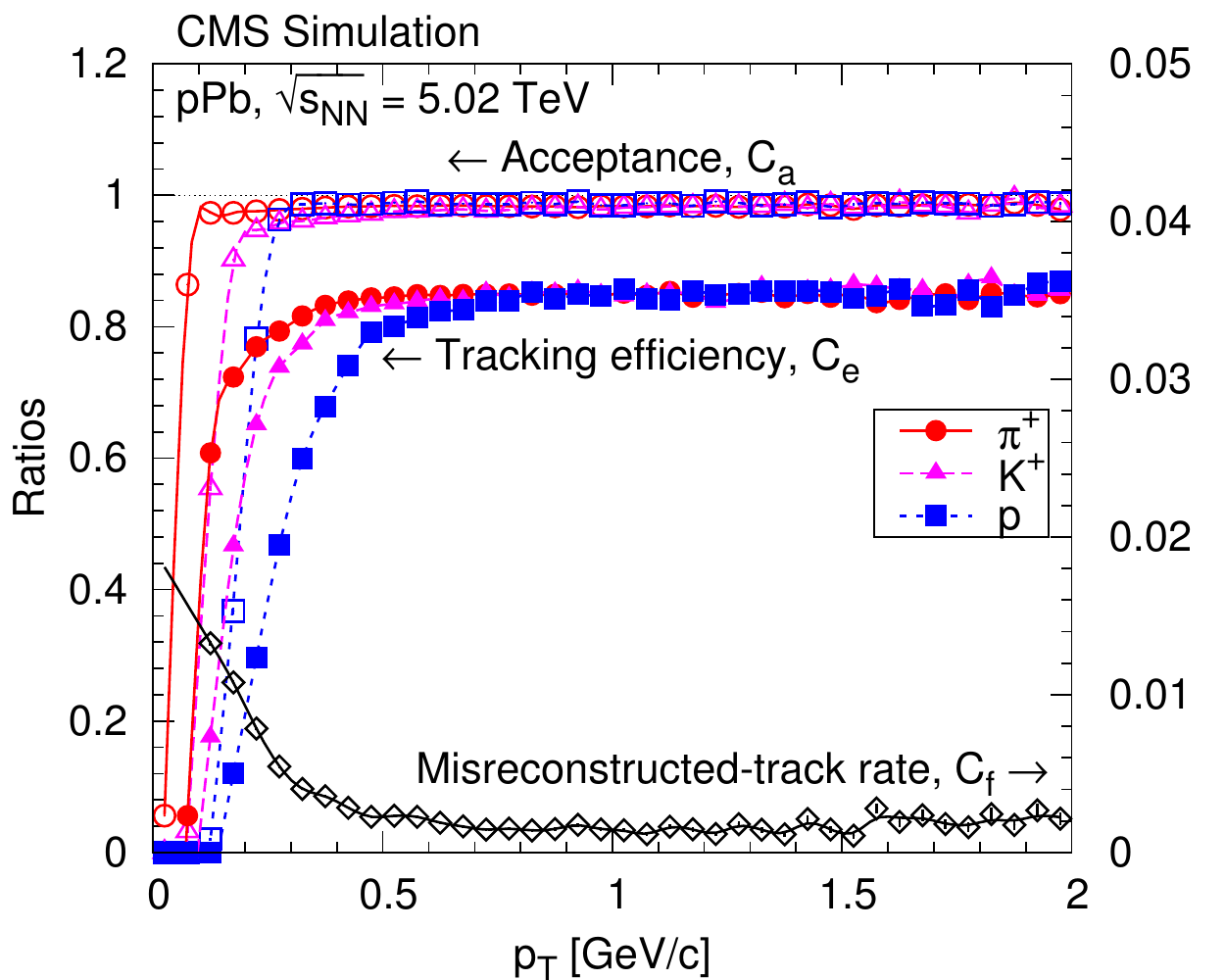}
 \end{center}

 \caption{\cmsLLeft: the ratio of selected events to double-sided (DS) events (ratio
of the corresponding efficiencies in the inelastic sample), according to
\EPOSLHC and \HIJING MC simulations, as a function of the reconstructed primary
charged-particle multiplicity.
\cmsRRight: acceptance, tracking efficiency (left scale), and misreconstructed-track
rate (right scale) in the range $\abs{\eta} < 2.4$ as a function of $\pt$ for
positively charged pions, kaons, and protons.}

 \label{fig:triggerEfficiency}

\end{figure}

The simulated ratio of the data selection efficiency to the DS selection
efficiency is shown as a function of the reconstructed track multiplicity in
the \cmsLeft panel of Fig.~\ref{fig:triggerEfficiency}. The ratio is used to
correct the measured events. The results are also corrected for the fraction of
DS events without a reconstructed track. This fraction, as given by the
simulation, is about 0.1\%.

The extrapolation of particle spectra into the unmeasured $(y,\pt)$ regions is
model dependent, particularly at low \pt. A high-precision measurement
therefore requires reliable track reconstruction down to the lowest possible
\pt. The present analysis extends to $\pt \approx 0.1\GeVc$ by exploiting
special tracking algorithms~\cite{Sikler:2007uh}, used in previous
studies~\cite{Khachatryan:2010xs,Khachatryan:2010us,identifiedSpectra}, to
provide high reconstruction efficiency and low background rate. The
charged-pion mass was assumed when fitting particle momenta.

The acceptance of the tracker ($C_\mathrm{a}$) is defined as the fraction of
primary charged particles leaving at least two hits in the pixel detector. It
is flat in the region $-2 < \eta < 2$ and $\pt > 0.4\GeVc$, and its value is
96--98\% (Fig.~\ref{fig:triggerEfficiency}, \cmsRight panel). The loss of
acceptance at $\pt < 0.4\GeVc$ is caused by energy loss and multiple scattering
of particles, both depending on the particle mass. Likewise, the reconstruction
efficiency ($C_\mathrm{e}$) is about 75--85\%, degrading at low \pt, also in a
mass-dependent way. The misreconstructed-track rate ($C_f$) is very small,
reaching 1\% only for $\pt <$ 0.2\GeVc. The probability of reconstructing
multiple tracks ($C_\mathrm{m}$) from a single true track is about 0.1\%,
mostly due to particles spiralling in the strong magnetic field of the CMS
solenoid. The efficiencies and background rates do not depend on the
charged-multiplicity of the event. They largely factorize in $\eta$ and \pt,
but for the final corrections an $(\eta,\pt)$ matrix is used.

The region where pPb collisions occur (beam spot) is measured by reconstructing
vertices from many events. Since the bunches are very narrow in the transverse
direction, the $xy$ location of the interaction vertices is well constrained;
conversely, their $z$ coordinates are spread over a relatively long distance
and must be determined on an event-by-event basis.
The vertex position is determined using reconstructed tracks which have $\pt >
0.1\GeVc$ and originate from the vicinity of the beam spot, i.e. their
transverse impact parameters $d_\mathrm{T}$ satisfy the condition $d_\mathrm{T}
< 3\,\sigma_T$.  Here $\sigma_\mathrm{T}$ is the quadratic sum of the
uncertainty in the value of $d_\mathrm{T}$ and the root-mean-square of the beam
spot distribution in the transverse plane.
The agglomerative vertex-reconstruction algorithm~\cite{Sikler:2009nx} was
used, with the $z$ coordinates (and their uncertainties) of the tracks at the
point of closest approach to the beam axis as input.
For single-vertex events, there is no minimum requirement on the number of
tracks associated with the vertex, even one-track vertices are allowed.  Only
tracks associated with a primary vertex are used in the analysis.  If multiple
vertices are present, the tracks from the highest multiplicity vertex are used.
The resultant bias is negligible since the pileup rate is extremely small.

The vertex reconstruction resolution in the $z$ direction is a strong function
of the number of reconstructed tracks and it is always smaller than
0.1\unit{cm}.
The distribution of the $z$ coordinates of the reconstructed primary vertices
is Gaussian, with a standard deviation of 7.1\unit{cm}. The simulated data were
reweighted so as to have the same vertex $z$ coordinate distribution as the
data.

The hadron spectra were corrected for particles of non-primary origin ($\tau >
10^{-12}\unit{s}$). The main sources of secondary particles are weakly decaying
particles, mostly \PKzS, \PgL/\PagL, and \PgSp/\PagSm. While the correction
($C_\mathrm{s}$) is around 1\% for pions, it rises up to 15\% for protons with
$\pt \approx 0.2\GeVc$. As none of the mentioned weakly decaying particles
decay into kaons, the correction for kaons is small.
Based on studies comparing reconstructed \PKzS, \PgL, and \PagL\ spectra and
predictions from the \HIJING\ event generator, the corrections are reweighted
by a \pt-dependent factor.

For $p < 0.15\GeVc$, electrons can be clearly identified. The overall $\Pe^\pm$
contamination of the hadron yields is below 0.2\%. Although muons cannot be
separated from pions, their fraction is very small, below 0.05\%. Since both
contaminations are negligible, no corrections are applied for them.

\section{Estimation of energy loss rate and yield extraction}

\label{sec:energyLoss}

In this paper an analytical parametrization~\cite{Sikler:2011yy} has been used
to approximate the energy loss of charged particles in the silicon detectors.
The method provides the probability density $P(\Delta|\varepsilon, l)$ of
energy deposit $\Delta$, if the most probable energy loss rate $\varepsilon$ at
a reference path-length $l_0 = 450\mum$ and the path-length $l$ are known. It
was used in conjunction with a maximum likelihood method, for the estimate of
$\varepsilon$.

For pixel clusters, the energy deposits were calculated as the sum of
individual pixel deposits.
In the case of strips, the energy deposits were corrected for capacitive
coupling and cross-talk between neighboring strips. The readout threshold, the
coupling parameter, and the standard deviation of the Gaussian noise for strips
were determined from data, using tracks with close-to-normal incidence.

For an accurate determination of $\varepsilon$, the response of all readout
chips was calibrated with multiplicative gain correction factors. The measured
energy deposit spectra were compared to the energy loss parametrization and
hit-level corrections (affine transformation of energy deposits using scale
factors and shifts) were introduced. The corrections were applied to individual
hits during the determination of the $\mpe$ fit templates (described below).

The best value of $\varepsilon$ for each track was calculated with the
corrected energy deposits by minimizing the joint energy deposit negative
log-likelihood of all hits on the trajectory (index $i$), $\chi^2 = -2 \sum_i
\ln P(\Delta_i|\varepsilon,l_i)$.  Hits with incompatible energy deposits
(contributing more than 12 to the joint $\chi^2$) were excluded. At most one
hit was removed; this affected about 1.5\% of the tracks.

\begin{figure}[htbp]

 \centering
  \includegraphics[width=0.49\textwidth]{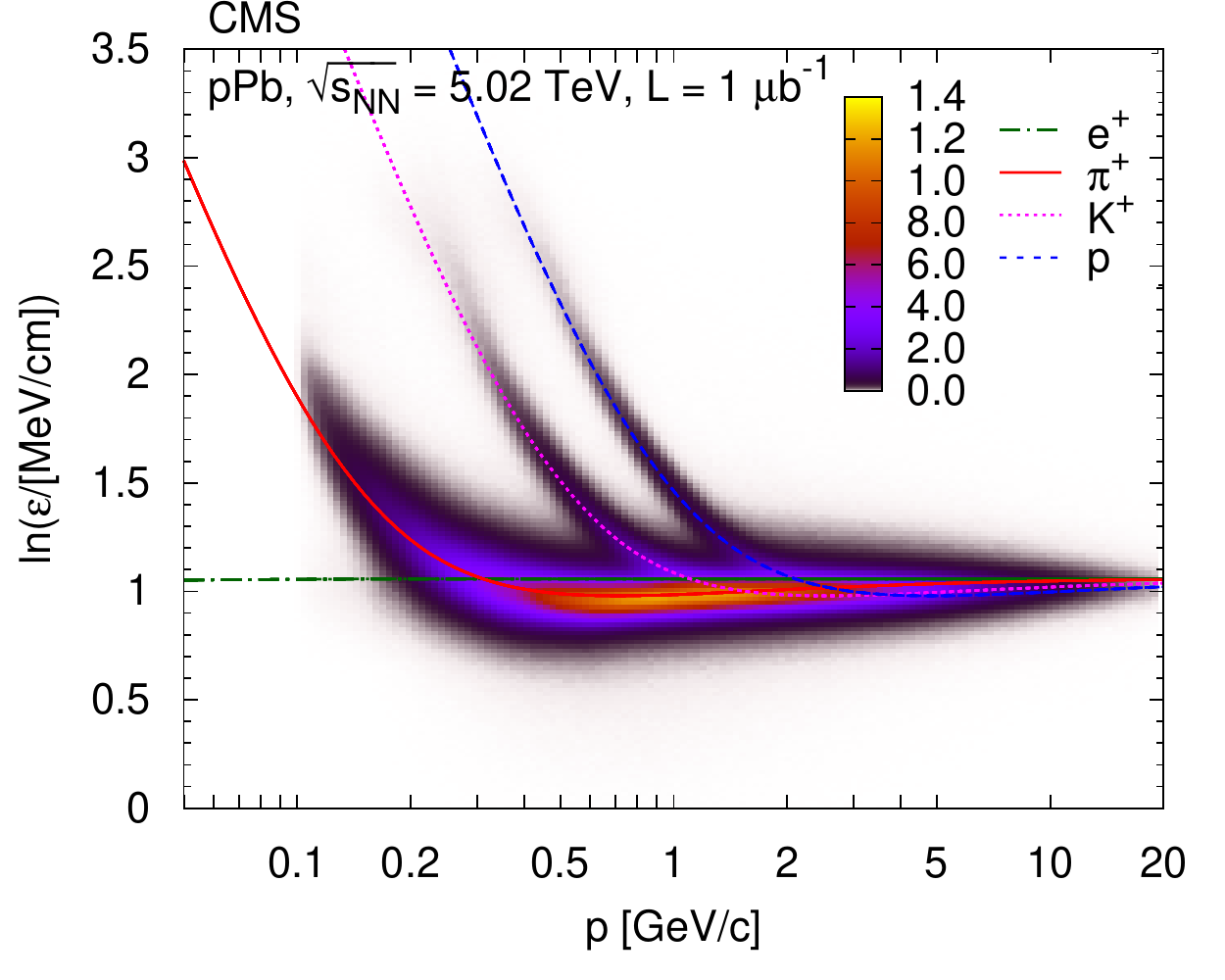}
  \includegraphics[width=0.49\textwidth]{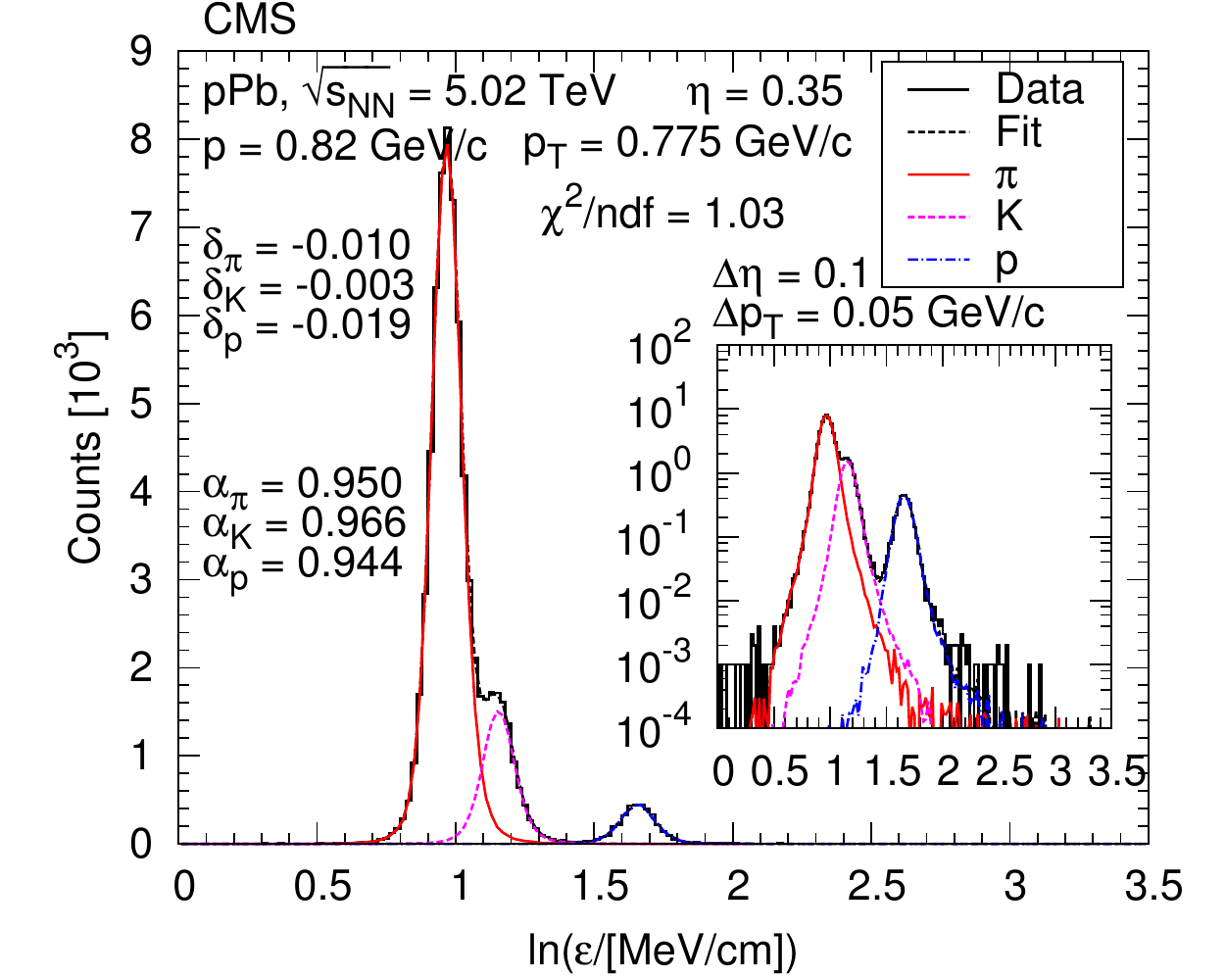}

 \caption{\cmsLLeft: distribution of $\mpe$ as a function of total momentum
$p$, for positively charged particles ($\varepsilon$ is the most probable
energy loss rate at a reference path length $l_0 = 450\mum$). The $z$ scale is
shown in arbitrary units and is linear. The curves show the expected $\mpe$ for
electrons, pions, kaons, and protons (Eq.~(30.11) in
Ref.~\cite{Beringer:1900zz}).
\cmsRRight: example $\mpe$ distribution at $\eta = 0.35$ and $\pt =
0.775\GeVc$, with bin widths $\Delta\eta = 0.1$ and $\Delta\pt = 0.05~\GeVc$.
Scale factors ($\alpha$) and shifts ($\delta$) are indicated (see text). The
inset shows the distribution with logarithmic vertical scale.}

 \label{fig:elossHistos}

\end{figure}

Distributions of $\mpe$ as a function of total momentum $p$ for positive
particles are plotted in the \cmsLeft panel of Fig.~\ref{fig:elossHistos} and
compared to the predictions of the energy loss method~\cite{Sikler:2011yy} for electrons, pions,
kaons, and protons.
The remaining deviations were taken into account by means of track-level
corrections mentioned above (affine transformation of templates using scale factors
and shifts, $\ln \varepsilon \rightarrow \alpha \ln \varepsilon + \delta$).

Low-momentum particles can be identified unambiguously and can therefore be
counted. Conversely, at high momentum, the $\mpe$ bands overlap (above about
0.5\GeVc for pions and kaons and 1.2\GeVc for protons); the particle yields
therefore need to be determined by means of a series of template fits in
$\mpe$, in bins of $\eta$ and \pt (Fig.~\ref{fig:elossHistos}, \cmsRight
panel).
Finally, fit templates, giving the expected $\mpe$ distributions for all
particle species (electrons, pions, kaons, and protons), were built from
tracks.  All kinematical parameters and hit-related observables were kept, but
the energy deposits were regenerated by sampling from the analytical
parametrization.
For a less biased determination of track-level residual corrections, enhanced
samples of each particle type were employed. These were used for setting
starting values of the fits.
For electrons and positrons, photon conversions in the beam-pipe and innermost
first pixel layer were used. For high-purity \Pgp\ and enhanced \Pp\ samples,
weakly decaying hadrons were selected (\PKzS, \PgL/\PagL).
The relations and constraints described in Ref.~\cite{identifiedSpectra} were
also exploited, this way better constraining the parameters of the fits:
 fitting the $\mpe$ distributions in number of hits ($\nh$) and track-fit
$\chi^2/\mathrm{ndf}$ slices simultaneously;
 fixing the distribution $\nh$ of particle species, relative to each other;
 using the expected continuity for refinement of track-level residual
corrections, in neighboring $(\eta,\pt)$ bins;
 using the expected convergence for track-level residual corrections, as the
$\mpe$ values of two particle species approach each other.

The results of the (iterative) $\mpe$ fits are the yields for each particle
species and charge in bins of $(\eta,\pt)$ or $(y,\pt)$, both inclusive and
divided into classes of reconstructed primary charged-track multiplicity.
In the end, the histogram fit $\chi^2/\mathrm{ndf}$ values were usually close
to unity.
Although pion and kaon yields could not be determined for $p > 1.30\GeVc$,
their sum was measured. This information is an important constraint when
fitting the \pt spectra.

The statistical uncertainties for the extracted yields are given by the fits.
The observed local variations of parameters in the $(\eta,\pt)$ plane for
track-level corrections cannot be attributed to statistical fluctuations and
indicate that the average systematic uncertainties in the scale factors and
shifts are about $10^{-2}$ and $2 \cdot 10^{-3}$, respectively.
These scale factors and shifts agree with those seen in the
high-purity samples to well within a factor of two.
The systematic uncertainties in the yields in each bin were obtained by
refitting the histograms with the parameters changed by these amounts.

\section{Corrections and systematic uncertainties}

\label{sec:corrections}

The measured yields in each $(\eta,\pt)$ bin, $\Delta N_\text{measured}$, were
first corrected for the misreconstructed-track rate ($C_\mathrm{f}$) and the
fraction of secondary particles ($C_\mathrm{s}$):

\begin{equation}
 \Delta N' = \Delta N_\text{measured}
  \cdot (1 - C_\mathrm{f}) \cdot (1 - C_\mathrm{s}).
\end{equation}

The distributions were then unfolded to take into account the finite $\eta$ and
\pt resolutions. The $\eta$ distribution of the tracks is almost flat and the
$\eta$ resolution is very good. Conversely, the \pt distribution is steep in
the low-momentum region and separate corrections in each $\eta$ bin were
necessary.
An unfolding procedure with linear regularization~\cite{Press:1058313} was
used, based on response matrices obtained from MC samples for each particle
species.

The corrected yields were obtained by applying corrections for acceptance
($C_\mathrm{a}$), efficiency ($C_\mathrm{e}$), and multiple track reconstruction rate ($C_\mathrm{m}$):

\begin{equation}
 \frac{1}{N_\text{ev}} \frac{\rd^2 N}{\rd\eta\, \rd\pt}_\text{corrected} =
  \frac{1}
       {C_\mathrm{a} \cdot C_\mathrm{e} \cdot (1 + C_\mathrm{m})}
       \frac{\Delta N'}{N_\text{ev} \Delta\eta \Delta\pt},
\end{equation}

\noindent where $N_\text{ev}$ is the corrected number of DS events
(Fig.~\ref{fig:triggerEfficiency}). Bins with acceptance smaller than 50\%,
efficiency smaller than 50\%, multiple-track rate greater than 10\%, or
containing less than 80 tracks were not used.

Finally, the differential yields $\rd^2N/\rd\eta\,\rd\pt$ were transformed to
invariant yields $\rd^2N/\rd y\,\rd \pt$ by multiplying with the Jacobian $E/p$
and the $(\eta,\pt)$ bins were mapped into a $(y,\pt)$ grid.
As expected, there is a small (5--10\%) $y$ dependence in the narrow region
considered ($\abs{y}<1$), depending on event multiplicity. The yields as a
function of $\pt$ were obtained by averaging over rapidity.

\begin{figure}[!tbhp]
\centering
  \includegraphics[width=0.49\textwidth]{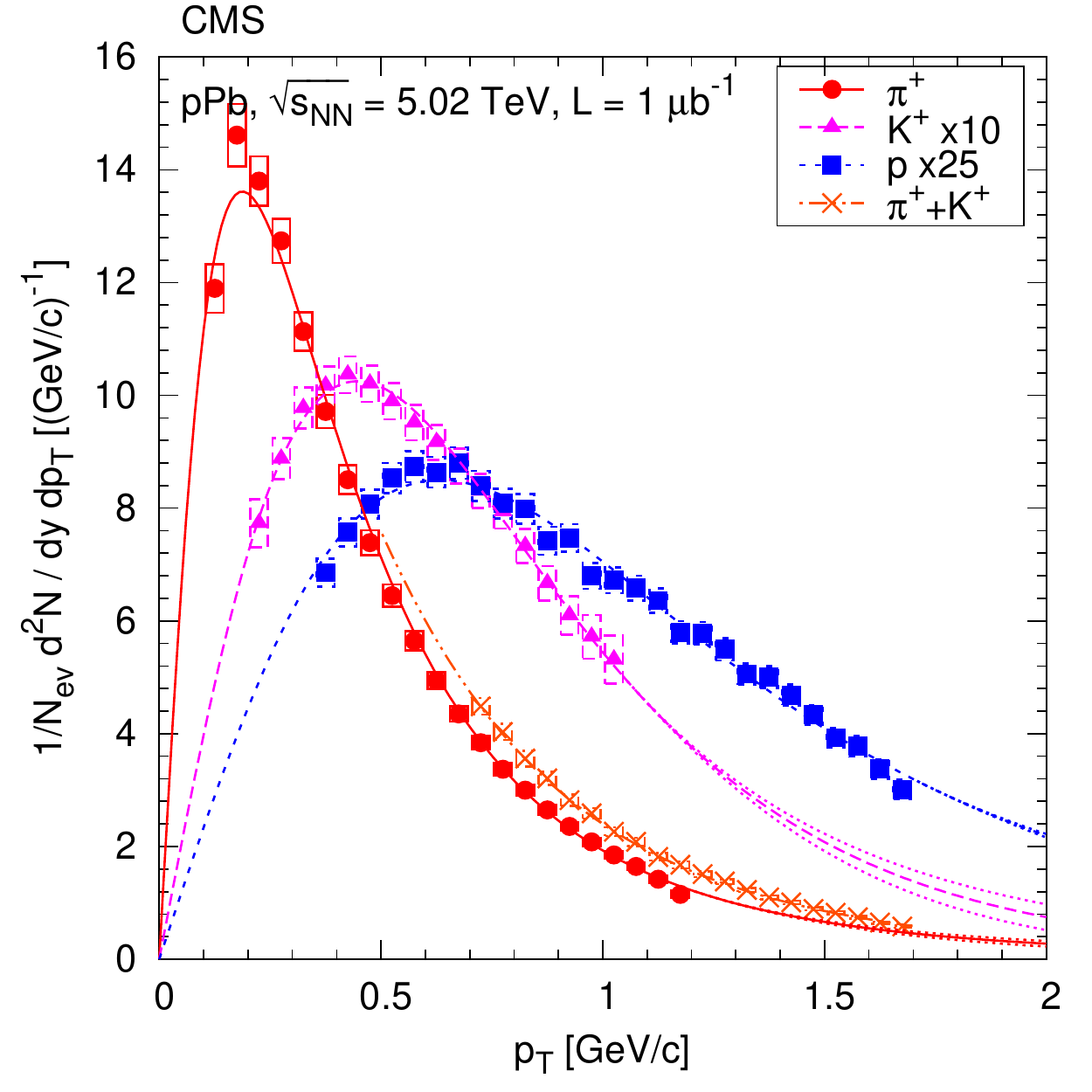}
  \includegraphics[width=0.49\textwidth]{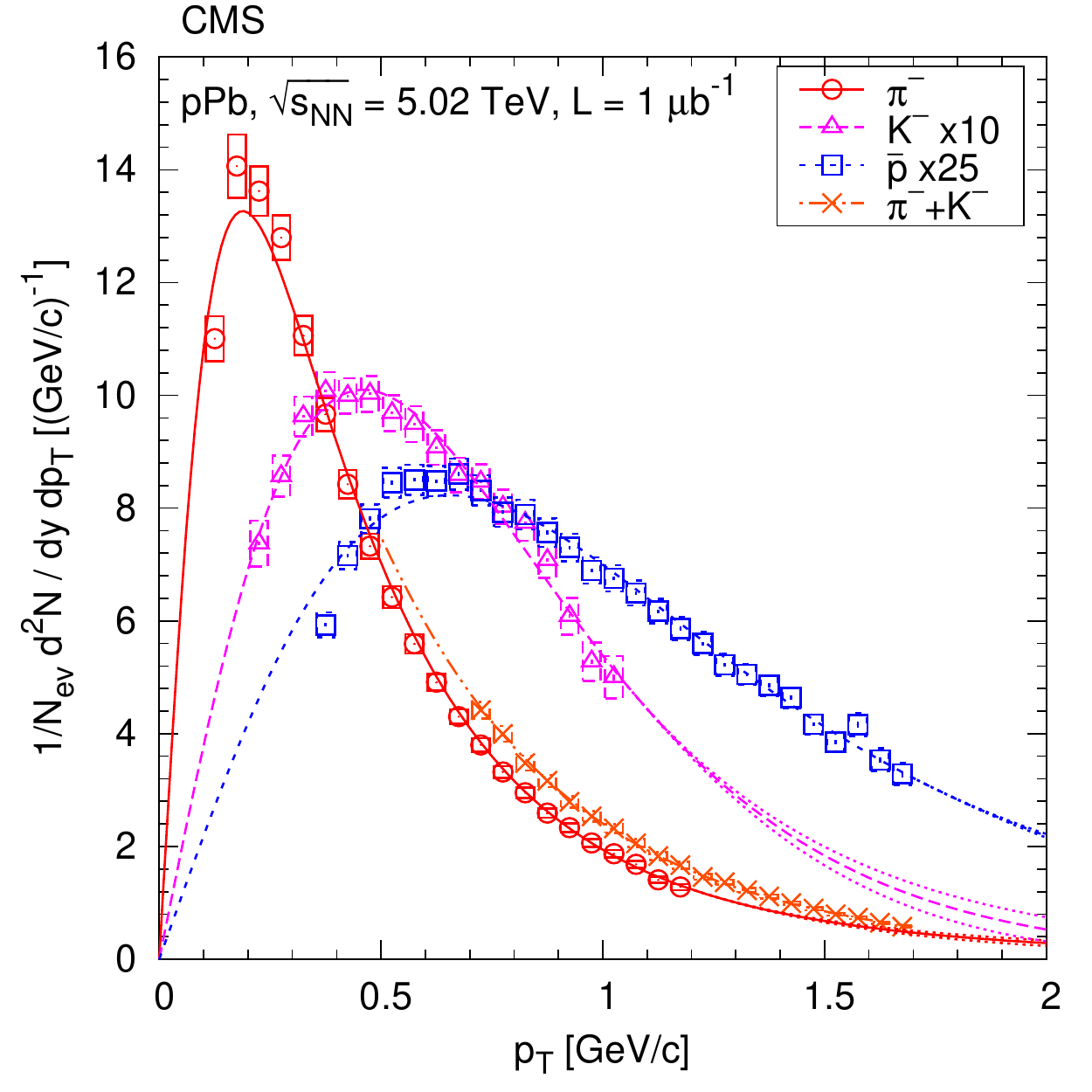}

 \caption{Transverse momentum distributions of identified charged hadrons
(pions, kaons, protons, sum of pions and kaons) in the range $\abs{y}<1$, for
positively (\cmsLeft) and negatively (\cmsRight) charged particles. Kaon and
proton distributions are scaled as shown in the legends. Fits to
Eqs.~\eqref{eq:tsallis} and \eqref{eq:tsallis2} are superimposed. Error bars
indicate the uncorrelated statistical uncertainties, while boxes show the
uncorrelated systematic uncertainties. The fully correlated normalization
uncertainty (not shown) is 3.0\%.
Dotted lines illustrate the effect of varying the $1/n$ value
of the Tsallis-Pareto function by $\pm 0.1$ above the highest measured \pt
point.}

 \label{fig:dndpt_lin}

\end{figure}

The systematic uncertainties are very similar to those in
Ref.~\cite{identifiedSpectra} and are summarized in Table~\ref{tab:error}.
The uncertainties of the corrections related to the event selection and pileup
are fully or mostly correlated and were treated as normalization uncertainties:
3.0\% uncertainty on the yields and 1.0\% on the average \pt.
In order to study the influence of the high \pt extrapolation
on $\langle \rd N/\rd y\rangle$ and $\langle \pt \rangle$, the $1/n$ parameter
of the fitted Tsallis-Pareto function (Sec.~\ref{sec:results}) was varied.
While keeping the function in the measured range, $1/n$ was increased and
decreased by $\pm 0.1$ above the highest \pt measured point, ensuring that the
two function pieces are continuous both in value and derivative. The choice of
the magnitude for the variation was motivated by the fitted $1/n$ values and
their distance from a Boltzmann distribution. (The resulting functions are
plotted in Fig.~\ref{fig:dndpt_lin} as dotted lines.) The high \pt
extrapolation introduces sizeable systematic uncertainties, 4--6\% for $\langle
\rd N/\rd y\rangle$, and 9--15\% for $\langle \pt \rangle$ in case of the DS
selection.

The tracker acceptance and the track reconstruction efficiency generally have
small uncertainties (1\% and 3\%, respectively), but change rapidly at very low
\pt (\cmsRight panel of Fig.~\ref{fig:triggerEfficiency}), leading to a 6\%
uncertainty on the yields in that range. For the multiple-track and
misreconstructed-track rate corrections, the uncertainty is assumed to be 50\%
of the correction, while for the case of the correction for secondary particles
it was estimated to be 20\%.
These mostly uncorrelated uncertainties are due to the imperfect modeling of
the detector: regions with mismodeled efficiency in the tracker, alignment
uncertainties, and channel-by-channel varying hit efficiency. These
circumstances can change frequently in momentum space, so can be treated as
uncorrelated.
 
The systematic uncertainties originating from the unfolding procedure were
studied. Since the $\pt$ response matrices are close to diagonal, the unfolding
of $\pt$ distributions did not introduce substantial systematics. At the same
time the inherited uncertainties were properly propagated. The introduced
correlations between neighboring $\pt$ bins were neglected, hence statistical
uncertainties were regarded as uncorrelated while systematic uncertainties were
expected to be locally correlated in $\pt$.
The systematic uncertainty of the fitted yields is in the range 1--10\%
depending mostly on total momentum.

\begin{table*}[htbp]

 \topcaption{Summary of the systematic uncertainties affecting the $\pt$
spectra. Values in parentheses indicate uncertainties in the
$\langle\pt\rangle$ measurement.
The systematic uncertainty related to the low \pt extrapolation is small
compared to the contributions from other sources and therefore not included in the
combined systematic uncertainty of the measurement.
Representative, particle-specific uncertainties (\Pgp, \PK, \Pp) are given for
$\pt =0.6\GeVc$ in the third group of systematic uncertainties.}

 \label{tab:error}

 \begin{center}
 \begin{tabular}{lcccc}
  \hline
  \multirow{2}{*}{\it Source} & \textit{Uncertainty} & \multicolumn{3}{c}{\textit{
Propagated}} \\
  & {\it of the source} [\%] & \multicolumn{3}{c}{\textit{
yield uncertainty} [\%]} \\
  \hline
  \multicolumn{3}{l}{Fully correlated, normalization} \\
  \; Correction for event selection & 3.0 (1.0)	& \multirow{3}{*}{$\biggl\}$} &
\multirow{3}{*}{4--6 (9--15)} & \multirow{3}{*}{} \\
  \; Pileup correction (merged and split vertices) & 0.3 & & & \\
  \; High \pt extrapolation  & 2--5 (8--15) & & & \\
  \hline
  \multicolumn{3}{l}{Mostly uncorrelated} \\
  \; Pixel hit efficiency		& 0.3 & \multirow{2}{*}{$\biggl\}$} &
\multirow{2}{*}{0.3} & \\
  \; Misalignment, different scenarios	& 0.1 & & & \\
  \hline
  \multicolumn{2}{l}{Mostly uncorrelated, $(y,\pt)$ dependent} & \Pgp & \PK & \Pp \\
  \; Acceptance of the tracker	        & 1--6 & 1 & 1 & 1 \\
  \; Efficiency of the reconstruction	& 3--6 & 3 & 3 & 3 \\
  \; Multiple-track reconstruction      & 50\% of the corr.
                                         & -- & -- & -- \\
  \; Misreconstructed-track rate	 & 50\% of the corr.
                                         & 0.1 & 0.1 & 0.1\\
  \; Correction for secondary particles 	& 20\% of the corr.
                                         & 0.2  & -- & 2 \\
  \; Fitting $\mpe$ distributions & 1--10 & 1    & 2    & 1  \\
  \hline
 \end{tabular}
 \end{center}

\end{table*}

\section{Results}

\label{sec:results}

In previously published measurements of unidentified and identified particle
spectra \cite{Khachatryan:2010xs,Khachatryan:2011tm}, the following form of the
Tsallis-Pareto-type distribution \cite{Tsallis:1987eu,Biro:2008hz} was fitted
to the data:

\begin{gather}
 \frac{\rd^2 N}{\rd y \, \rd\pt} =
  \frac{\rd N}{\rd y} \cdot C
                \cdot \pt \left[1 + \frac{\mt - m}{nT} \right]^{-n},
 \label{eq:tsallis}
\intertext{where}
 C = \frac{(n-1)(n-2)}{nT[nT + (n-2) m]}
\end{gather}

\noindent and $\mt = \sqrt{m^2 + \pt^2}$
(factors of $c$ are omitted from the preceding formulae).
The free parameters are the integrated yield $\rd N/\rd y$, the exponent $n$,
and parameter $T$.
The above formula is useful for extrapolating the spectra to zero \pt and very
high $\pt$ and for extracting $\langle\pt\rangle$ and $\rd N/\rd y$. Its
validity for different multiplicity bins was cross-checked by fitting MC
spectra in the $\pt$ ranges where there are data points, and verifying that the
fitted values of $\langle\pt\rangle$ and $\rd N/\rd y$ were consistent with the
generated values.
Nevertheless, for a more robust estimation of both quantities
($\langle\pt\rangle$ and $\langle \rd N/\rd y \rangle$), the data points and
their uncertainties were used in the measured range and the fitted functions
only for the extrapolation in the unmeasured regions.
According to some models of particle production based on non-extensive
thermodynamics~\cite{Biro:2008hz}, the parameter $T$ is connected with the
average particle energy, while $n$ characterizes the ``non-extensivity'' of the
process, i.e. the departure of the spectra from a Boltzmann distribution ($n =
\infty$).

As discussed earlier, pions and kaons cannot be unambiguously distinguished at
higher momenta. Because of this, the pion-only, the kaon-only, and the joint
pion and kaon $\rd^2N/\rd y \, \rd\pt$ distributions were fitted
for $\abs{y} < 1$ and $p < 1.20\GeVc$,
    $\abs{y} < 1$ and $p < 1.05\GeVc$,
    and $\abs{\eta} <1$ and $1.05 < p < 1.7\GeVc$, respectively.
Since the ratio $p/E$ for the pions (which are more abundant than
kaons) at these momenta can be approximated by $\pt/\mt$ at $\eta \approx 0$,
Eq.~\eqref{eq:tsallis} becomes:

\begin{equation}
 \frac{\rd^2 N}{\rd\eta\, \rd\pt} \approx
  \frac{\rd N}{\rd y} \cdot C
                \cdot \frac{\pt^2}{\mt} \left(1 + \frac{\mt-m}{nT} \right)^{-n}.
 \label{eq:tsallis2}
\end{equation}

\begin{figure}[!htbp]
\centering

  \includegraphics[width=0.49\textwidth]{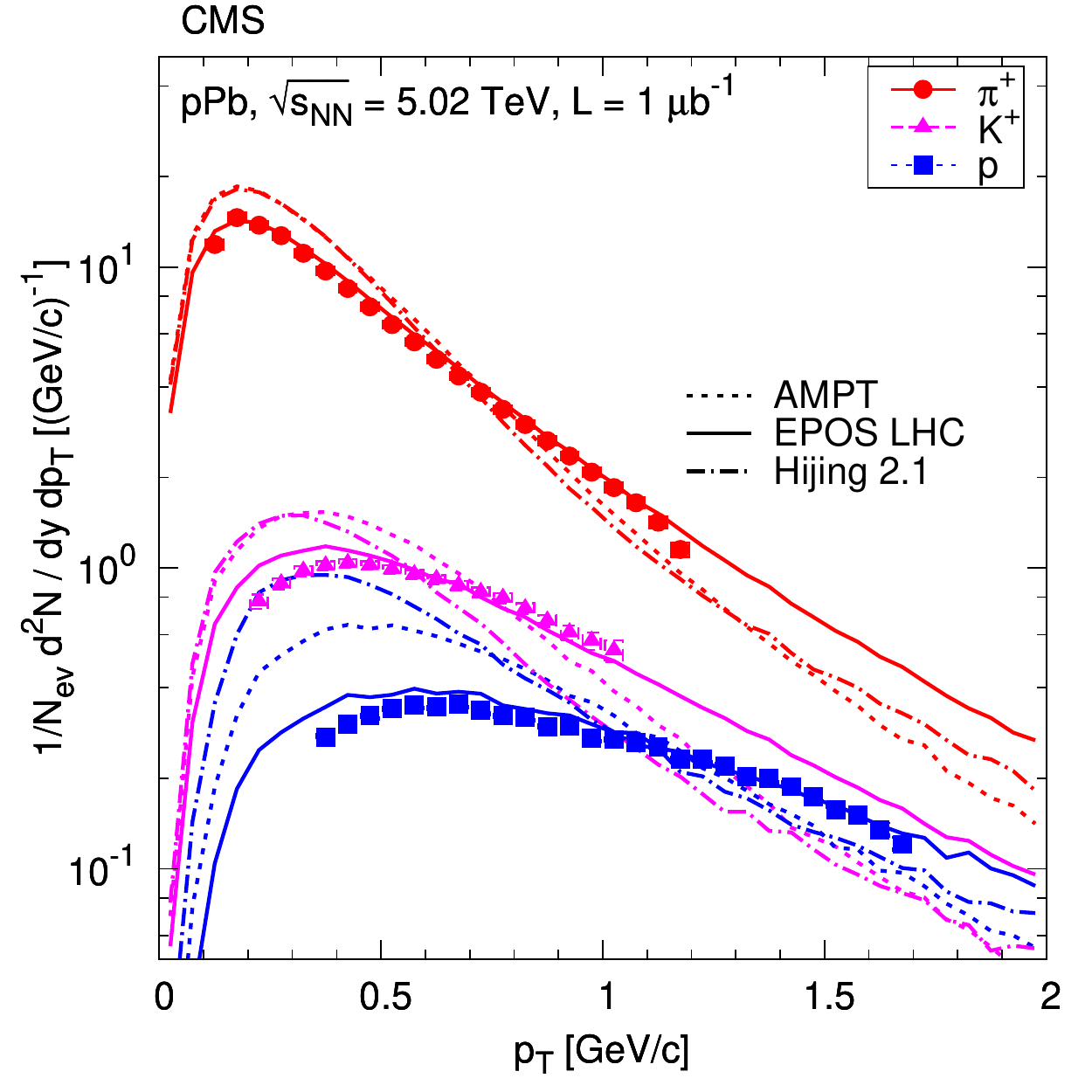}
  \includegraphics[width=0.49\textwidth]{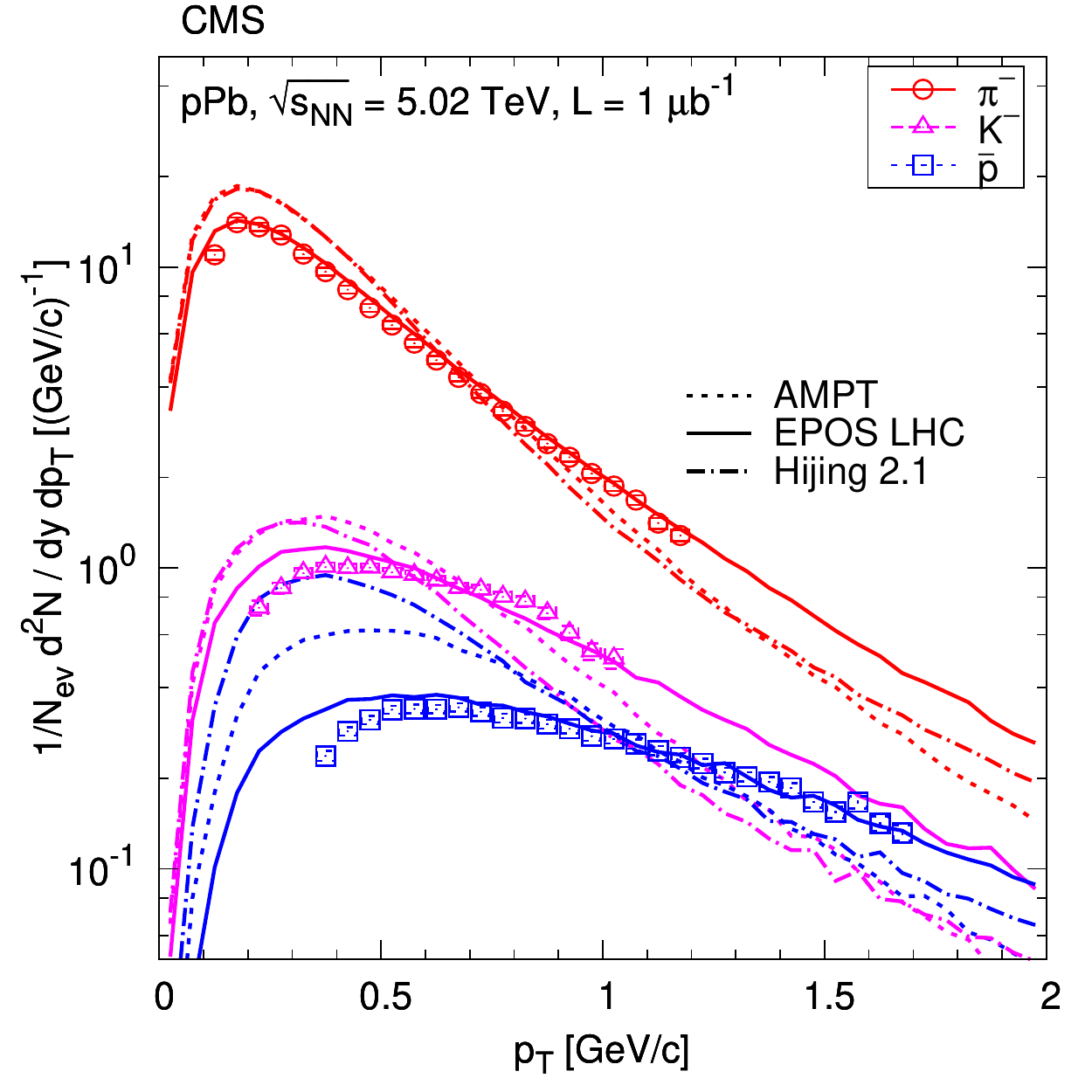}

 \caption{Transverse momentum distributions of identified charged hadrons
(pions, kaons, protons) in the range $\abs{y}<1$, for positively (\cmsLeft) and
negatively (\cmsRight) charged particles. Measured values (same as in
Fig.~\ref{fig:dndpt_lin}) are plotted together with predictions from \AMPT,
\EPOSLHC, and \HIJING. Error bars indicate the uncorrelated statistical
uncertainties, while boxes show the uncorrelated systematic uncertainties. The
fully correlated normalization uncertainty (not shown) is 3.0\%.}

 \label{fig:dndpt_log}

\end{figure}

The approximate fractions of particles outside the measured \pt range depend on
track multiplicity; they are 15--30\% for pions, 40--50\% for kaons, and
20--35\% for protons.
The average transverse momentum $\langle \pt \rangle$ and its uncertainty were
obtained using data points in the measured range complemented by numerical
integration of Eq.~\eqref{eq:tsallis} with the fitted parameters in the
unmeasured regions, under the assumption that the particle yield distributions
follow the Tsallis-Pareto function in the low-$\pt$ and high-$\pt$ regions.

The results discussed in the following are for laboratory rapidity $\abs{y} <
1$. In all cases, error bars indicate the uncorrelated statistical
uncertainties, while boxes show the uncorrelated systematic uncertainties. The
fully correlated normalization uncertainty is not shown. For the \pt spectra,
the average transverse momentum, and the ratio of particle yields, the data are
compared to \AMPT 1.26/2.26~\cite{Lin:2011zzg},
\EPOSLHC~\cite{Werner:2005jf,Pierog:2013ria}, and
\HIJING 2.1~\cite{Deng:2010mv} MC event generators.
Numerical results corresponding to the plotted spectra, fit results, as well as
their statistical and systematic uncertainties are given in Ref.~\cite{data}.

\subsection{Inclusive measurements}

The transverse momentum distributions of positively and negatively charged
hadrons (pions, kaons, protons) are shown in Fig.~\ref{fig:dndpt_lin}, along
with the results of the fits to the Tsallis-Pareto parametrization
(Eqs.~\eqref{eq:tsallis} and \eqref{eq:tsallis2}). The fits are of good quality
with $\chi^2/\mathrm{ndf}$ values in the range 0.4--2.8 (Table~\ref{tab:fits}).
Figure~\ref{fig:dndpt_log} presents the data compared to the \AMPT, \EPOSLHC,
and \HIJING predictions. \EPOSLHC gives a good description, while other
generators predict steeper \pt distributions than found in data.

\begin{table*}[thbp]

 \topcaption{Fit results ($\rd N/\rd y$, $1/n$, and $T$) and goodness-of-fit
values for the DS selection shown together with calculated averages
($\langle\rd N/\rd y\rangle$, $\langle\pt\rangle$) for charged pions, kaons,
and protons.
The systematic uncertainty related to the low \pt extrapolation is small
compared to the contributions from other sources and therefore not included in
the combined systematic uncertainty of the measurement.
Combined uncertainties are given.
}

 \label{tab:fits}

 \begin{center}
 \begin{tabular}{ccrcccc}
 \hline
  {Particle} & $\rd N/\rd y$ & \multicolumn{1}{c}{$1/n$} & $T$ [$\GeVcns{}$]
             & $\chi^2/\mathrm{ndf}$
             & $\langle \rd N/\rd y \rangle$
             & $\langle \pt \rangle$ [$\GeVcns{}$] \\
 \hline
 \Pgpp & 8.074 $\pm$ 0.081 & 0.190 $\pm$ 0.007 & 0.131 $\pm$ 0.003
       & 0.88
       & 8.064 $\pm$ 0.190    
       & 0.547 $\pm$ 0.078 \\ 
 \Pgpm & 7.971 $\pm$ 0.079 & 0.195 $\pm$ 0.007 & 0.131 $\pm$ 0.003
       & 1.05
       & 7.966 $\pm$ 0.196
       & 0.559 $\pm$ 0.083 \\
  \PKp & 1.071 $\pm$ 0.068 & 0.092 $\pm$ 0.066 & 0.278 $\pm$ 0.022
       & 0.42
       & 1.040 $\pm$ 0.053
       & 0.790 $\pm$ 0.104 \\
  \PKm & 0.984 $\pm$ 0.047 & $-$0.008 $\pm$ 0.067 & 0.316 $\pm$ 0.024
       & 2.82
       & 0.990 $\pm$ 0.037
       & 0.744 $\pm$ 0.061 \\
  \Pp  & 0.510 $\pm$ 0.018 & 0.151 $\pm$ 0.036 & 0.325 $\pm$ 0.016
       & 0.81
       & 0.510 $\pm$ 0.024
       & 1.243 $\pm$ 0.183 \\
  \Pap & 0.494 $\pm$ 0.017 & 0.123 $\pm$ 0.038 & 0.349 $\pm$ 0.017
       & 1.32
       & 0.495 $\pm$ 0.022
       & 1.215 $\pm$ 0.165 \\
 \hline
 \end{tabular}
 \end{center}

\end{table*}

\begin{figure}[!htbp]

 \centering

  \includegraphics[width=0.49\textwidth]{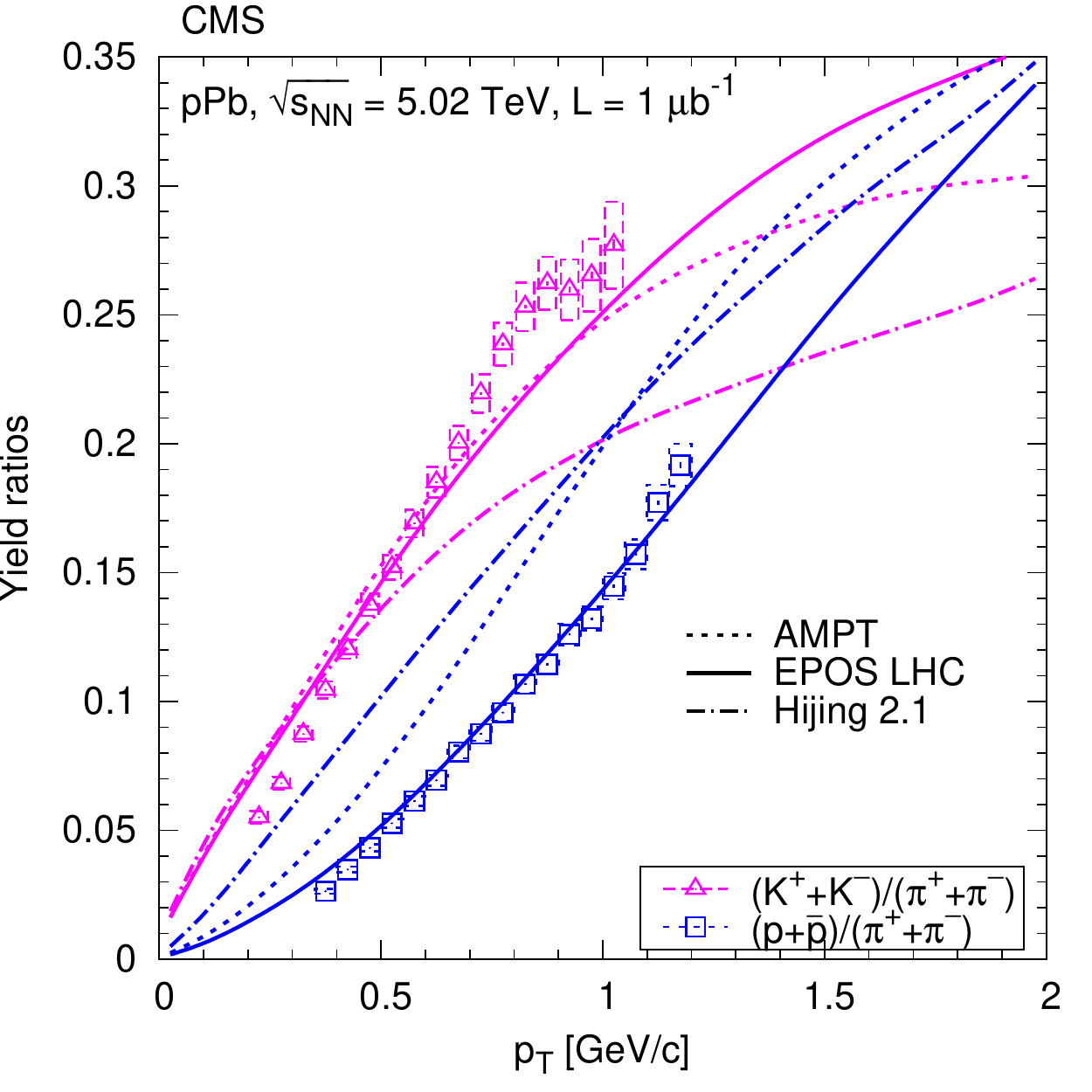}
  \includegraphics[width=0.49\textwidth]{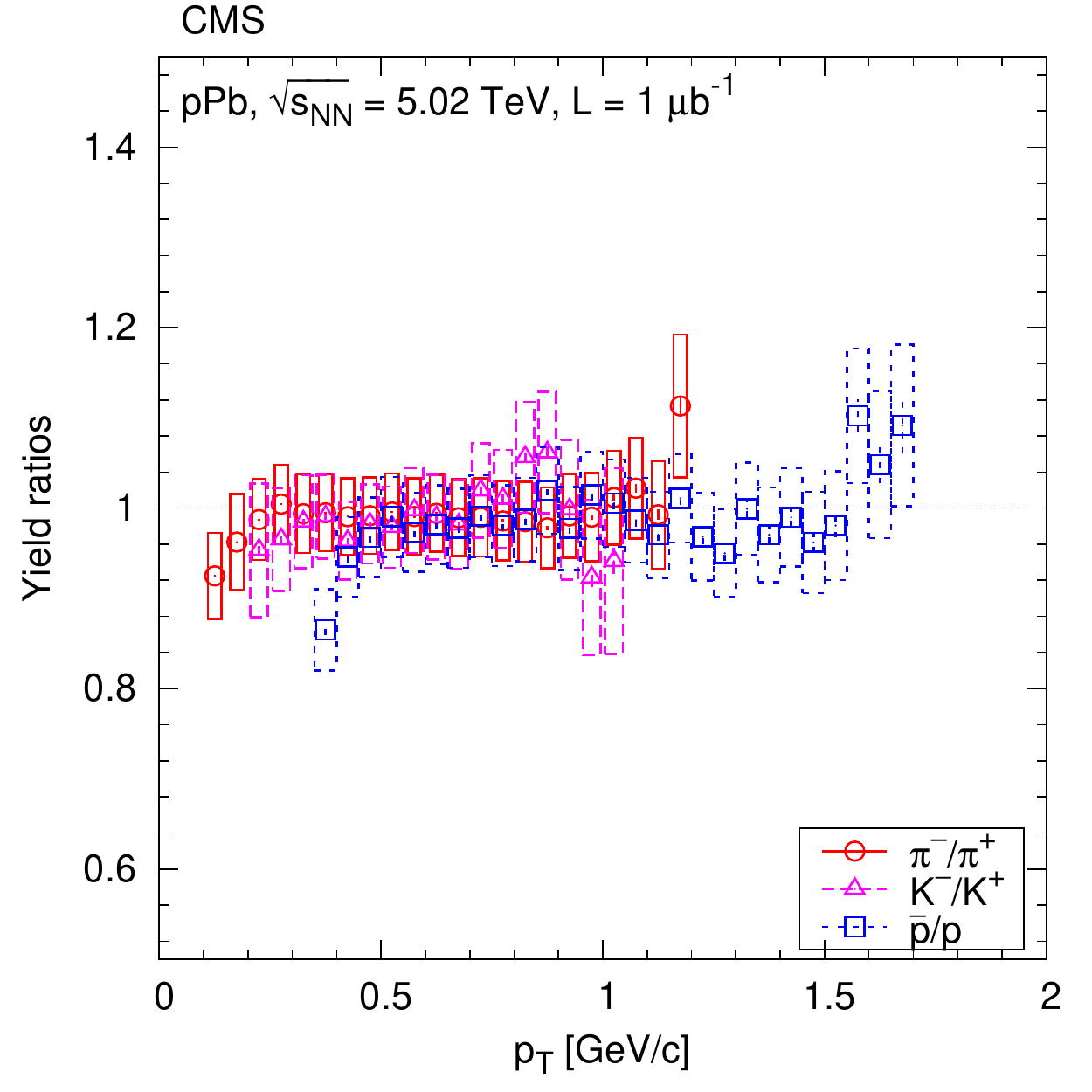}

 \caption{Ratios of particle yields as a function of transverse momentum.
\PK/\Pgp\ and \Pp/\Pgp\ values are shown in the \cmsLeft panel, and
opposite-charge ratios are plotted in the \cmsRight panel.
Error bars indicate the uncorrelated statistical uncertainties, while boxes
show the uncorrelated systematic uncertainties. In the \cmsLeft panel, curves
indicate predictions from \AMPT, \EPOSLHC, and \HIJING.}

 \label{fig:ratios_vs_pt}

\end{figure}

Ratios of particle yields as a function of the transverse momentum are plotted
in Fig.~\ref{fig:ratios_vs_pt}. While the $\PK/\Pgp$ ratios are well described
by the \AMPT simulation, only \EPOSLHC is able to predict both $\PK/\Pgp$ and
$\Pp/\Pgp$ ratios.
The ratios of the yields for oppositely charged particles are close to one, as
expected for LHC energies at midrapidity.

\subsection{Multiplicity dependent measurements}

A study of the dependence on track multiplicity is motivated partly by the
intriguing hadron correlations measured in pp and pPb collisions at high track
multiplicities~\cite{Khachatryan:2010gv,CMS:2012qk,Abelev:2012ola,Aad:2012gla},
suggesting possible collective effects in ``central'' pp and pPb collisions at
the LHC\@.
At the same time, it was seen that in pp collisions the characteristics of
particle production ($\langle \pt \rangle$, ratios) at LHC energies are
strongly correlated with event particle multiplicity rather than with the
center-of-mass energy of the collision~\cite{identifiedSpectra}.  The strong
dependence on multiplicity (or centrality) was also seen in dAu collisions at
RHIC~\cite{Abelev:2008ab,Adare:2013esx}.
In addition, the multiplicity dependence of particle yield ratios is sensitive
to various final-state effects (hadronization, color reconnection, collective
flow) implemented in MC models used in collider and cosmic-ray
physics~\cite{d'Enterria:2011kw}.

\begin{figure*}[!htb]

 \centering
  \includegraphics[width=\cmsFigWidth]{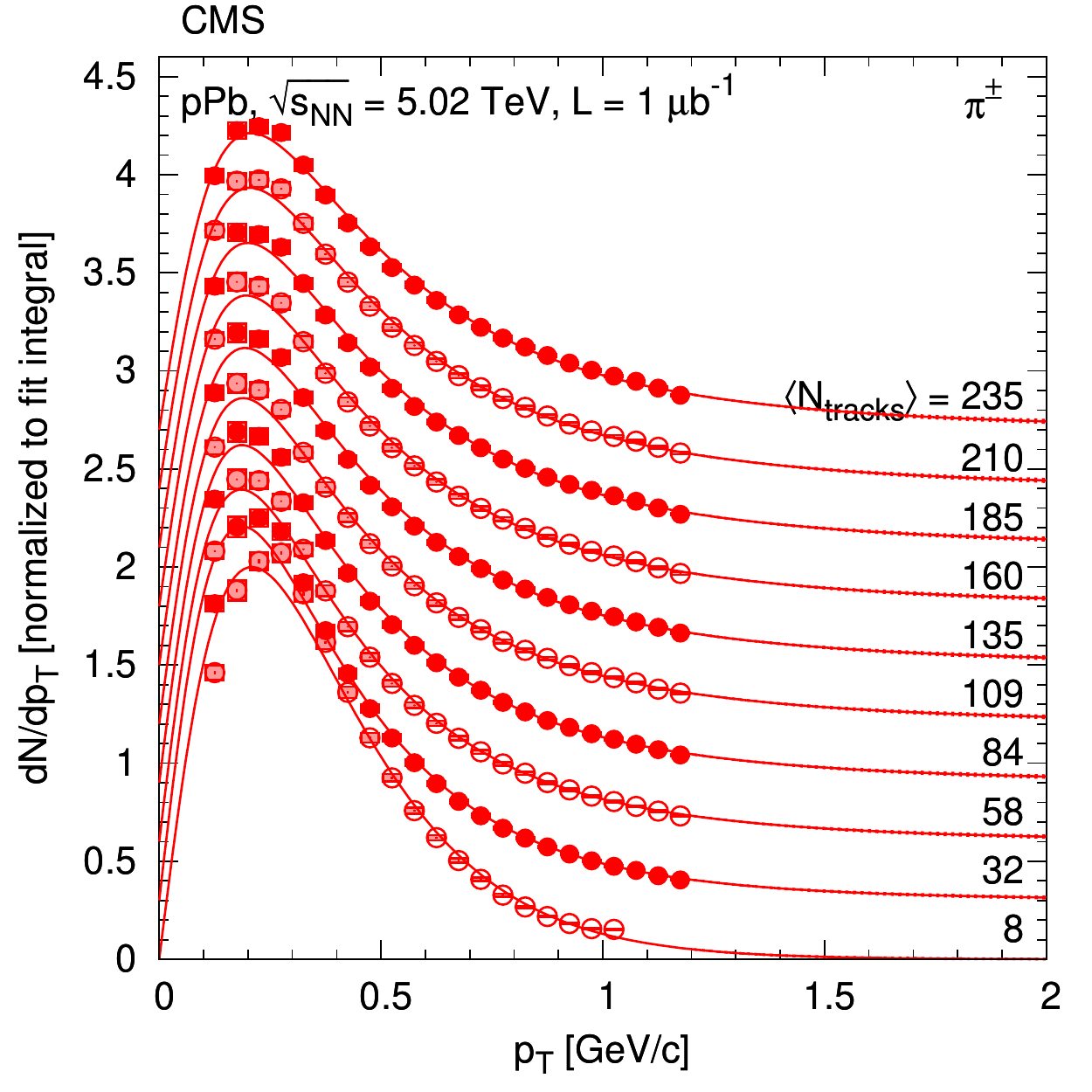}
  \includegraphics[width=\cmsFigWidth]{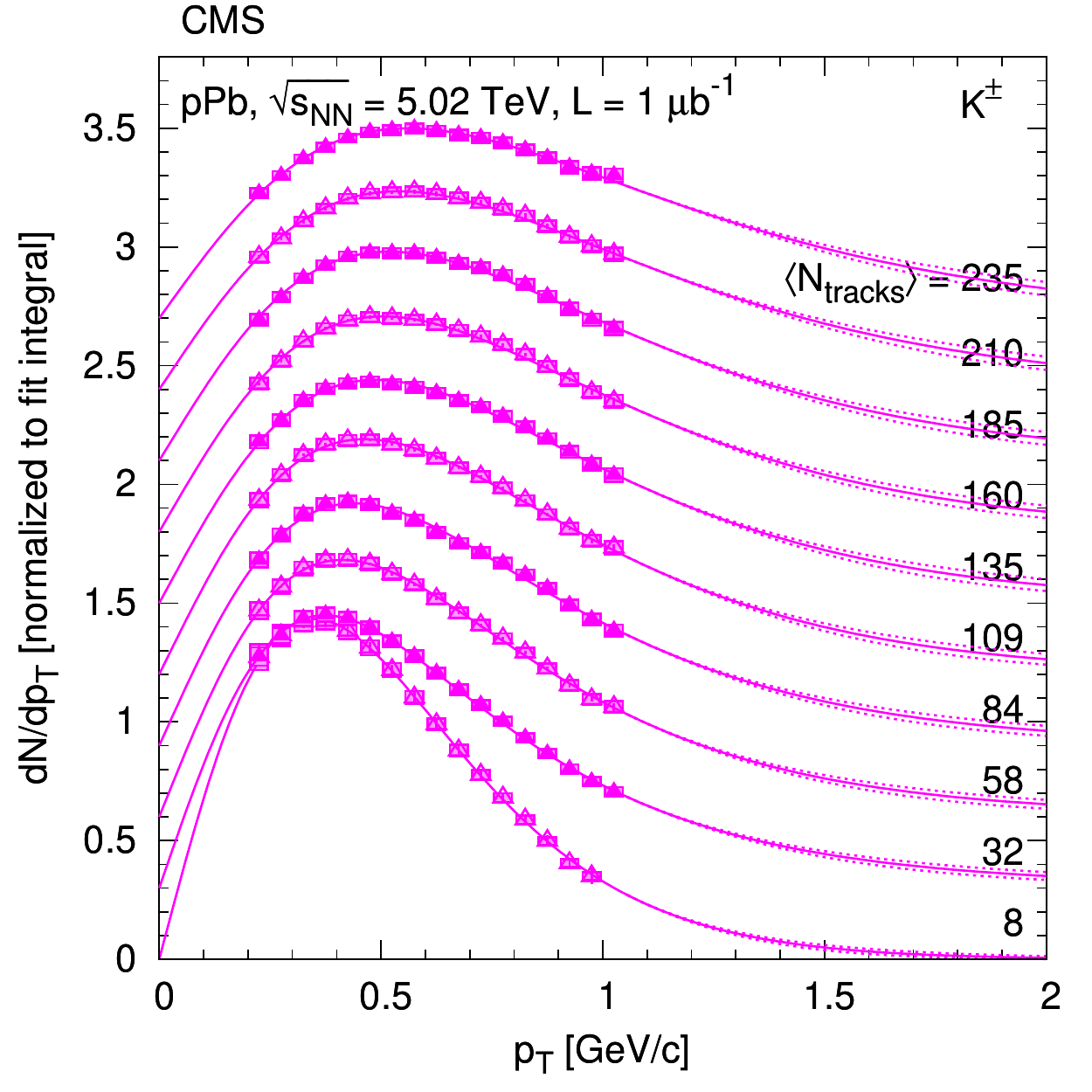}
  \includegraphics[width=\cmsFigWidth]{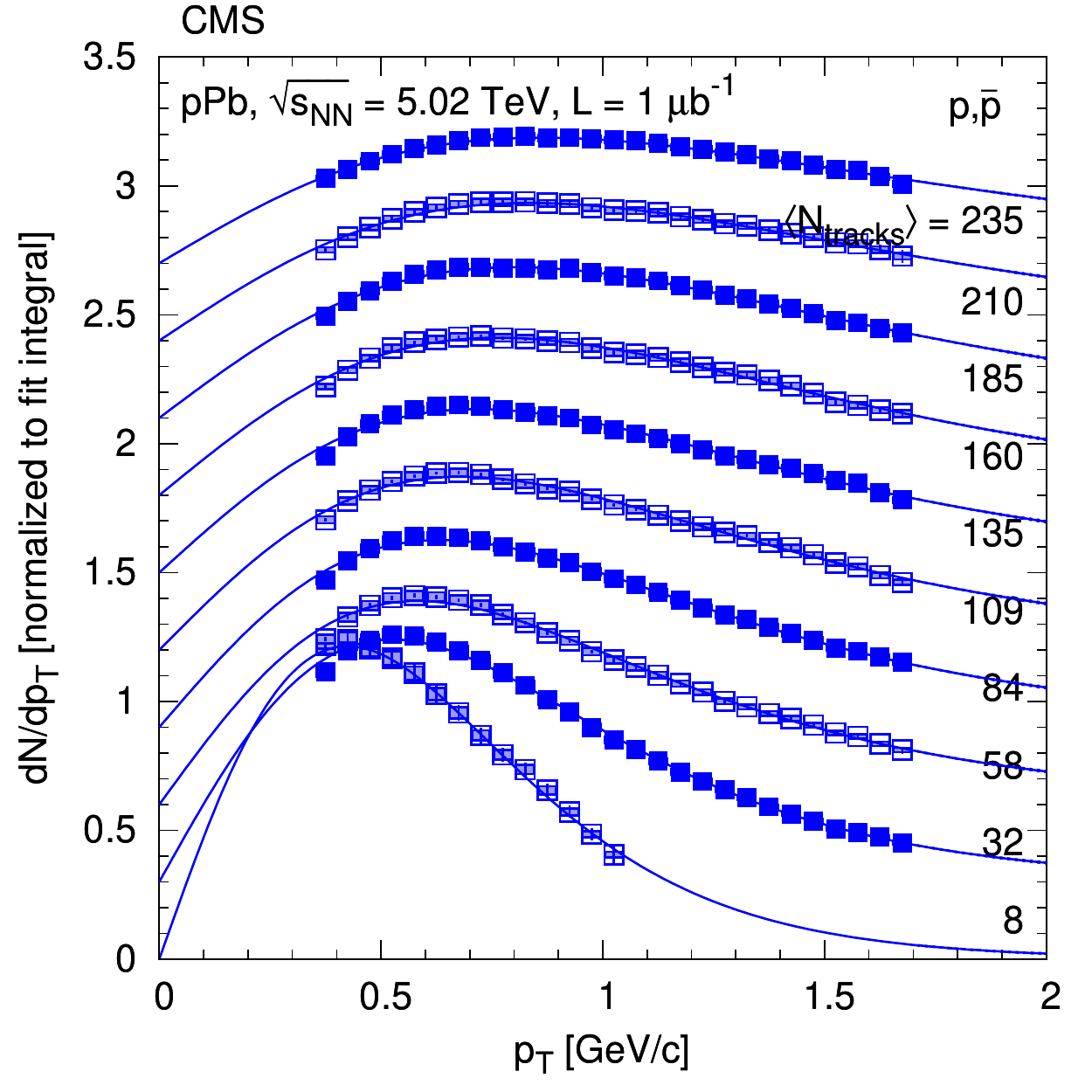}

 \caption{Transverse momentum distributions of charged pions, kaons, and
protons, normalized such that the fit integral is unity, in every second
multiplicity class ($\langle N_\text{tracks} \rangle$ values are indicated) in
the range $\abs{y}<1$, fitted with the Tsallis-Pareto parametrization (solid
lines).
For better visibility, the result for any given $\langle N_\text{tracks}
\rangle$ bin is shifted by 0.3 units with respect to the adjacent bins.
Error bars indicate the uncorrelated statistical uncertainties, while boxes
show the uncorrelated systematic uncertainties.
Dotted lines illustrate the effect of varying the $1/n$ value
of the Tsallis-Pareto function by $\pm 0.1$ above the highest measured \pt
point.
}

 \label{fig:dndpt_multi}

\end{figure*}

\begin{table*}[!bhtp]

 \topcaption{Relationship between the number of reconstructed tracks
($N_\text{rec}$) and the average number of corrected tracks ($\langle
N_\text{tracks} \rangle$) in the region $\abs{\eta} < 2.4$, and also with the
condition $\pt > 0.4\GeVc$~(used in Ref.~\cite{CMS:2012qk}), in the 19
multiplicity classes considered.}

 \label{tab:multiClass}

 \begin{center}
 \begin{tabular}{cc@{\hspace{\sptw}}c@{\hspace{\sptw}}c@{\hspace{\sptw}}c@{\hspace{\sptw}}c@{\hspace{\sptw}}c@{\hspace{\sptw}}c@{\hspace{\sptw}}c@{\hspace{\sptw}}c@{\hspace{\sptw}}c@{\hspace{\sptw}}c@{\hspace{\sptw}}c@{\hspace{\sptw}}c@{\hspace{\sptw}}c@{\hspace{\sptw}}c@{\hspace{\sptw}}c@{\hspace{\sptw}}c@{\hspace{\sptw}}c@{\hspace{\sptw}}c}
  \hline\\[-1.5ex]
  $N_\text{rec}$ &
  \begin{sideways}   0--9   \end{sideways} &
  \begin{sideways}  10--19  \end{sideways} &
  \begin{sideways}  20--29  \end{sideways} &
  \begin{sideways}  30--39  \end{sideways} &
  \begin{sideways}  40--49  \end{sideways} &
  \begin{sideways}  50--59  \end{sideways} &
  \begin{sideways}  60--69  \end{sideways} &
  \begin{sideways}  70--79  \end{sideways} &
  \begin{sideways}  80--89  \end{sideways} &
  \begin{sideways}  90--99  \end{sideways} &
  \begin{sideways} 100--109 \end{sideways} &
  \begin{sideways} 110--119 \end{sideways} &
  \begin{sideways} 120--129 \end{sideways} &
  \begin{sideways} 130--139 \end{sideways} &
  \begin{sideways} 140--149 \end{sideways} &
  \begin{sideways} 150--159 \end{sideways} &
  \begin{sideways} 160--169 \end{sideways} &
  \begin{sideways} 170--179 \end{sideways} &
  \begin{sideways} 180--189 \end{sideways} \\
  \hline
  $\langle N_\text{tracks} \rangle$ &
     8 &  19 &  32 &  45 &  58 &  71 &  84 &  96 & 109 & 122 &
   135 & 147 & 160 & 173 & 185 & 198 & 210 & 222 & 235 \\
  $\langle N_\text{tracks} \rangle_{\pt > 0.4\GeVc}$ &
    3 &   8 &  15 & 22 &  29 &
   36 &  43 &  50 & 58 &  65 &
   73 &  80 &  87 & 95 & 103 &
  110 & 117 & 125 & 133 \\
  \hline
 \end{tabular}
 \end{center}

\end{table*}

\begin{figure}[!thbp]

 \centering

  \includegraphics[width=0.49\textwidth]{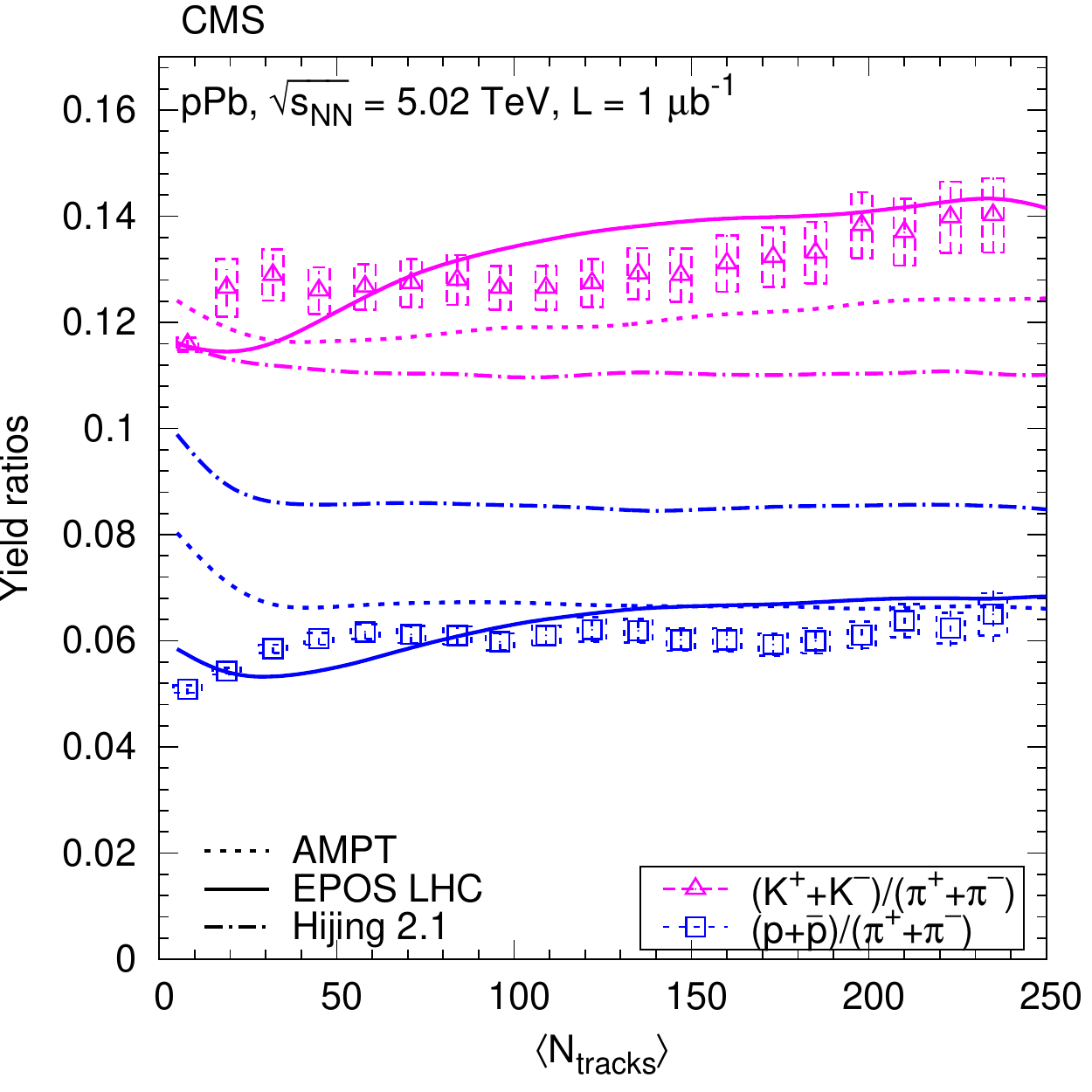}
  \includegraphics[width=0.49\textwidth]{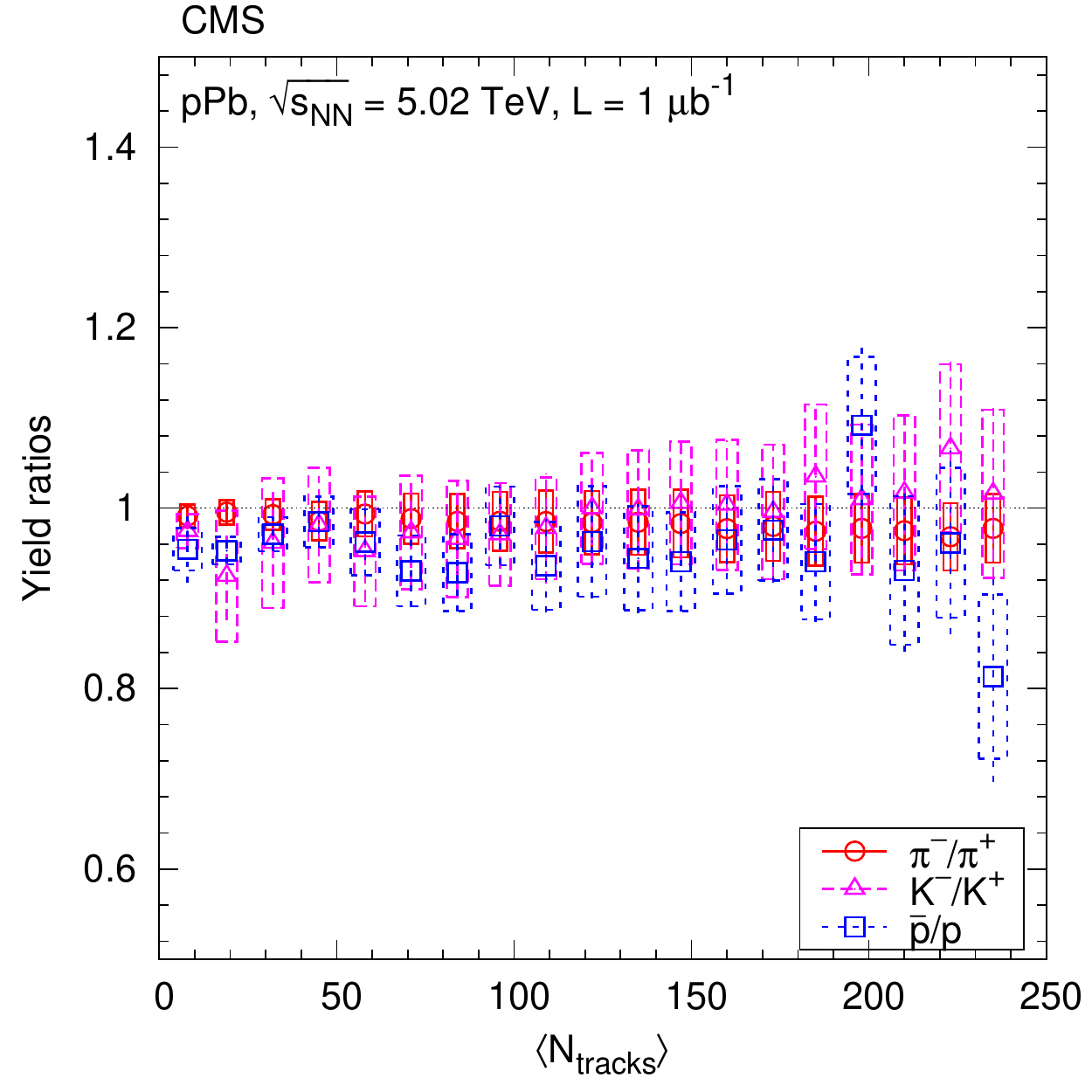}

 \caption{Ratios of particle yields in the range $\abs{y}<1$ as a function of
the corrected track multiplicity for $\abs{\eta}<2.4$.
\PK/\Pgp\ and \Pp/\Pgp\ values are shown in the \cmsLeft panel, and
opposite-charge ratios are plotted in the \cmsRight panel.
Error bars indicate the uncorrelated combined uncertainties, while boxes show
the uncorrelated systematic uncertainties. In the \cmsLeft panel, curves
indicate predictions from \AMPT, \EPOSLHC, and \HIJING.}

 \label{fig:ratios_vs_multi}

\end{figure}

\begin{figure}[!thbp]
 \centering
  \includegraphics[width=0.49\textwidth]{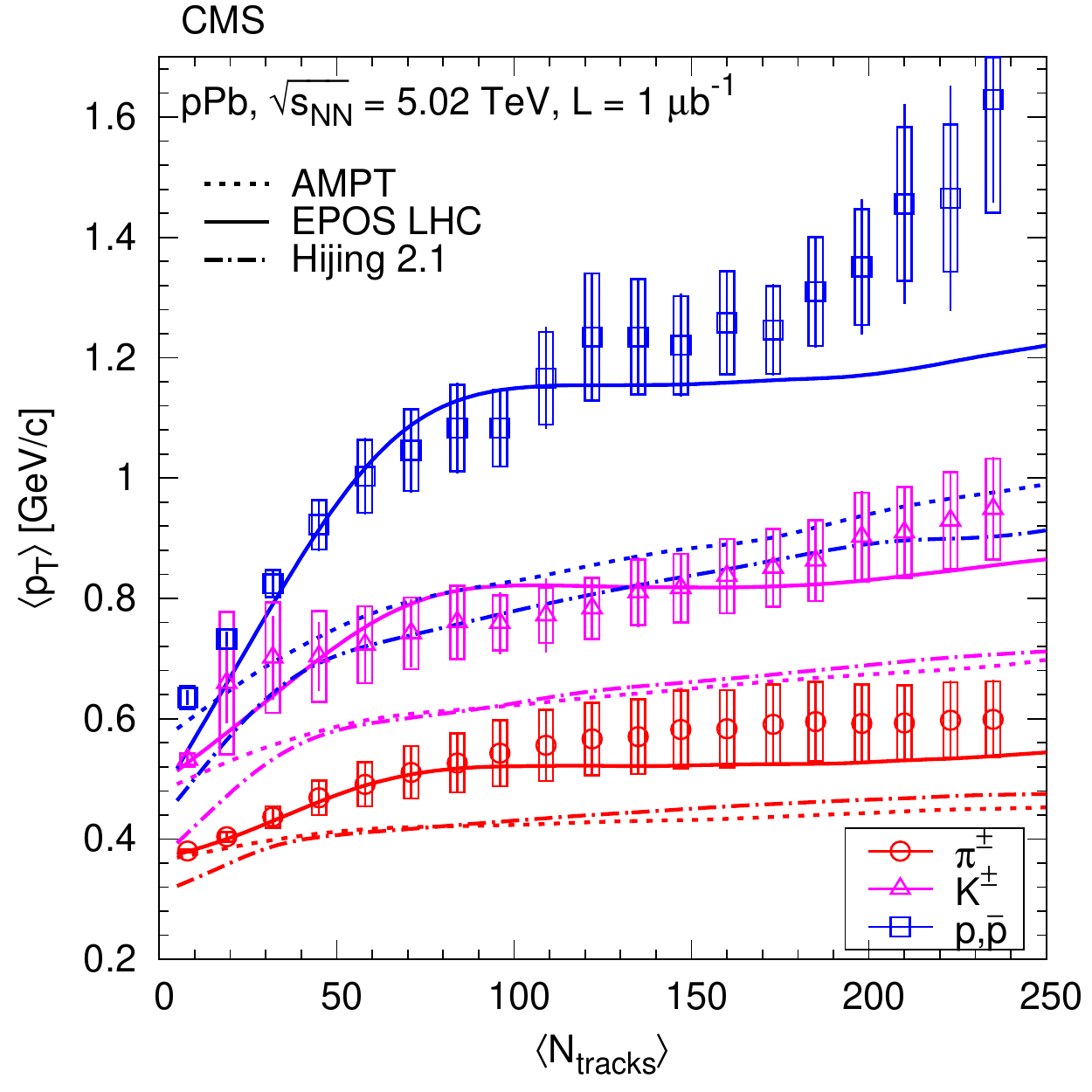}

 \caption{Average transverse momentum of identified charged hadrons (pions,
kaons, protons) in the range $\abs{y}<1$, as a function of the corrected track
multiplicity for $\abs{\eta}<2.4$, computed assuming a Tsallis-Pareto
distribution in the unmeasured range. Error bars indicate the uncorrelated
combined uncertainties, while boxes show the uncorrelated systematic
uncertainties. The fully correlated normalization uncertainty (not shown) is
1.0\%. Curves indicate predictions from \AMPT, \EPOSLHC, and \HIJING.}

 \label{fig:apt_vs_multi}

\end{figure}

\begin{figure}[!thbp]

 \centering
  \includegraphics[width=0.49\textwidth]{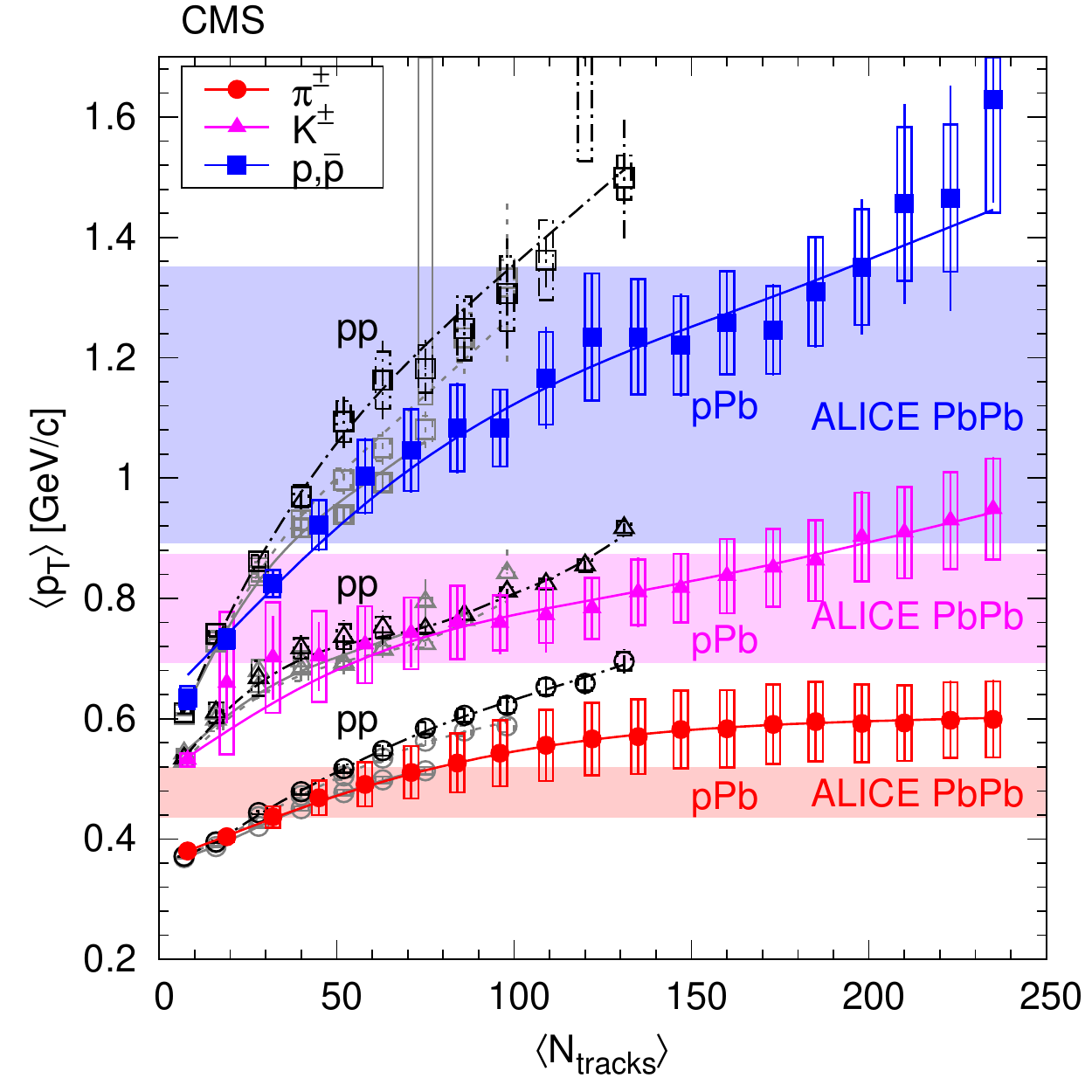}
  \includegraphics[width=0.49\textwidth]{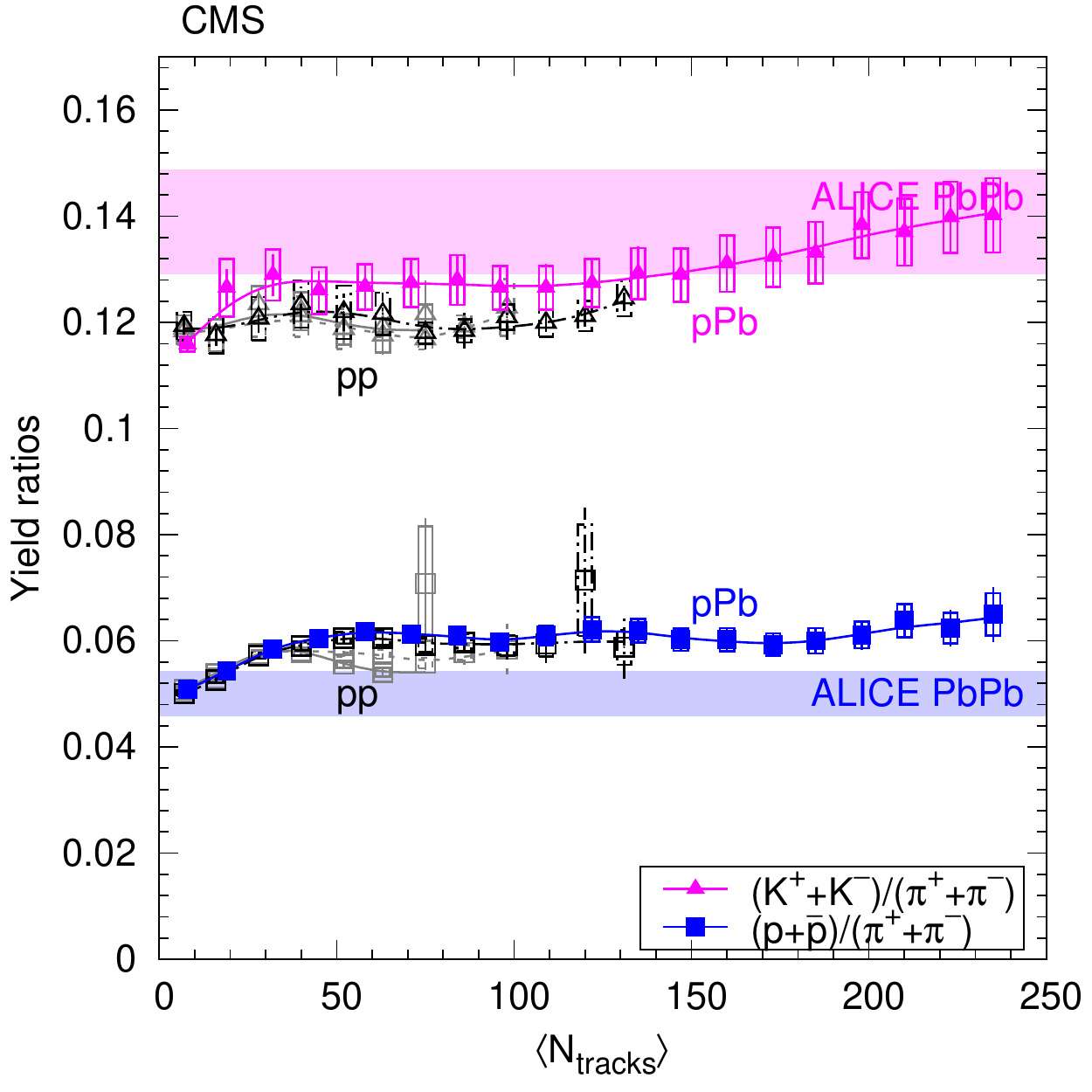}

 \caption{Average transverse momentum of identified charged hadrons (pions,
kaons, protons; \cmsLeft panel) and ratios of particle yields (\cmsRight panel)
in the range $\abs{y}<1$ as a function of the corrected track multiplicity for
$\abs{\eta}<2.4$, for pp collisions (open symbols) at several
energies~\cite{identifiedSpectra}, and for pPb collisions (filled symbols) at
$\sqrt{s_{NN}} =$ 5.02\TeV.
Both $\langle\pt\rangle$ and yield ratios were computed assuming a
Tsallis-Pareto distribution in the unmeasured range.
Error bars indicate the uncorrelated combined uncertainties, while boxes show
the uncorrelated systematic uncertainties. For $\langle\pt\rangle$ the fully
correlated normalization uncertainty (not shown) is 1.0\%.
In both plots, lines are drawn to guide the eye (gray solid -- pp 0.9\TeV, gray
dotted -- pp 2.76\TeV, black dash-dotted -- pp 7\TeV, colored solid -- pPb
5.02\TeV).
The ranges of $\langle\pt\rangle$, $\PK/\Pgp$ and $\Pp/\Pgp$ values measured by
ALICE in various centrality PbPb collisions (see text) at $\sqrt{s_{NN}} =
2.76\TeV$~\cite{Abelev:2013vea} are indicated with horizontal bands.}

 \label{fig:multiplicityDependence}

\end{figure}

The event multiplicity $N_\text{rec}$ is obtained from the number of
reconstructed tracks with $\abs{\eta}<2.4$, where the tracks are reconstructed
using the same algorithm as for the identified charged
hadrons~\cite{Sikler:2007uh}. (The multiplicity variable
$N_\text{trk}^\text{offline}$, used in Ref.~\cite{CMS:2012qk}, is obtained from
a different track reconstruction configuration and a value of
$N_\text{trk}^\text{offline} = 110$ corresponds roughly to $N_\text{rec} =
170$.) The event multiplicity was divided into 19 classes, defined in
Table~\ref{tab:multiClass}.
To facilitate comparisons with models, the corresponding corrected charged
particle multiplicity in the same acceptance of $\abs{\eta}<2.4$
($N_\text{tracks}$) is also determined.
For each multiplicity class, the correction from $N_\text{rec}$ to
$N_\text{tracks}$ uses the efficiency estimated with the \HIJING simulation in
$(\eta,\pt)$ bins. The corrected data are then integrated over \pt, down to
zero yield at $\pt = 0$ (with a linear extrapolation below $\pt = 0.1~\GeVc$).
Finally, the integrals for each eta slice are summed.
The average corrected charged-particle multiplicity $\langle N_\text{tracks}
\rangle$, and also its values with the condition $\pt > 0.4\GeVc$, are shown in
Table~\ref{tab:multiClass} for each event multiplicity class. The value of
$\langle N_\text{tracks} \rangle$ is used to identify the multiplicity class in
Figs.~\ref{fig:dndpt_multi}--\ref{fig:invslope}.

Transverse-momentum distributions of identified charged hadrons, normalized
such that the fit integral is unity,
in selected multiplicity classes for $\abs{y} < 1$ are shown in
Fig.~\ref{fig:dndpt_multi} for pions, kaons, and protons. The distributions of
negatively and positively charged particles have been summed. The distributions
are fitted with the Tsallis-Pareto parametrization
with $\chi^2/\mathrm{ndf}$ values in the range
0.8--4.0 for pions,
0.1--1.1 for kaons, and
0.1--0.7 for protons.
For kaons and protons, the parameter $T$ increases with multiplicity, while for
pions $T$ slightly increases and the exponent $n$ slightly decreases with
multiplicity (not shown).

The ratios of particle yields are displayed as a function
of track multiplicity in Fig.~\ref{fig:ratios_vs_multi}. The $\PK/\Pgp$ and
$\Pp/\Pgp$ ratios are flat, or slightly rising, as a function of
$\langle N_\text{tracks} \rangle$.
While none of the models is able to precisely reproduce the track multiplicity
dependence, the best and worst matches to the overall scale are given by
\EPOSLHC and \HIJING, respectively.
The ratios of yields of oppositely charged particles are independent of
$\langle N_\text{tracks} \rangle$ as shown in the \cmsRight panel of Fig.~\ref{fig:ratios_vs_multi}.
The average transverse momentum $\langle\pt\rangle$ is shown as a function of
multiplicity in Fig.~\ref{fig:apt_vs_multi}. As expected from the discrepancies
between theory and data shown in Fig.~\ref{fig:dndpt_log}, \EPOSLHC again gives
a reasonable description, while the other event generators presented here
underpredict the measured values.
For the dependence of $T$ on multiplicity (not shown), the predictions match
the pion data well; the kaon and proton values are much higher than in \AMPT or
\HIJING.

\subsection{Comparisons to pp and PbPb data}

\label{sec:comparisons}

The comparison with pp data taken at various center-of-mass energies (0.9,
2.76, and 7\TeV)~\cite{identifiedSpectra} is shown in
Fig.~\ref{fig:multiplicityDependence}, where the dependence of
$\langle\pt\rangle$ and the particle yield ratios ($\PK/\Pgp$ and $\Pp/\Pgp$)
on the track multiplicity is shown. The plots also display the ranges of these
values measured by ALICE in PbPb collisions at $\sqrt{s_{NN}} =$ 2.76\TeV for
centralities from peripheral (80--90\% of the inelastic cross-section) to
central (0--5\%)~\cite{Abelev:2013vea}. These ALICE PbPb data cover a much
wider range of $N_\text{tracks}$ than is shown in the plot.
Although PbPb data are not available at $\sqrt{s_{NN}} = 5.02\TeV$ for
comparison, the evolution of event characteristics from RHIC ($\sqrt{s_{NN}} =
0.2\TeV$,~\cite{Adler:2003cb,Arsene:2005mr,Abelev:2008ab}) to LHC energies
\cite{Abelev:2013vea} suggests that yield ratios should remain similar, while
$\langle\pt\rangle$ values will increase by about 5\% when going from
$\sqrt{s_{NN}} =$ 2.76\TeV to 5.02\TeV.

\begin{figure}[htbp]

 \centering
  \includegraphics[width=0.49\textwidth]{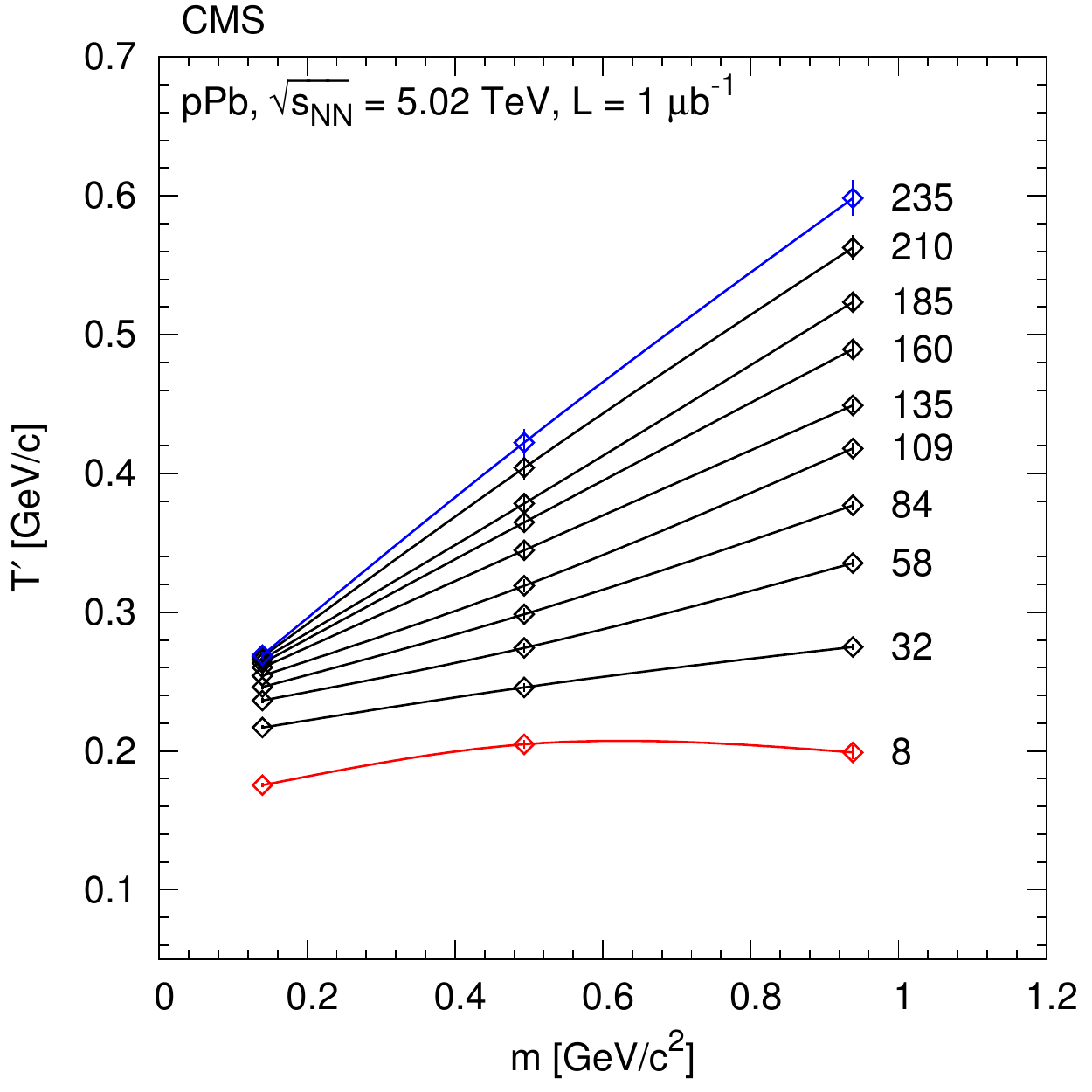}
  \includegraphics[width=0.49\textwidth]{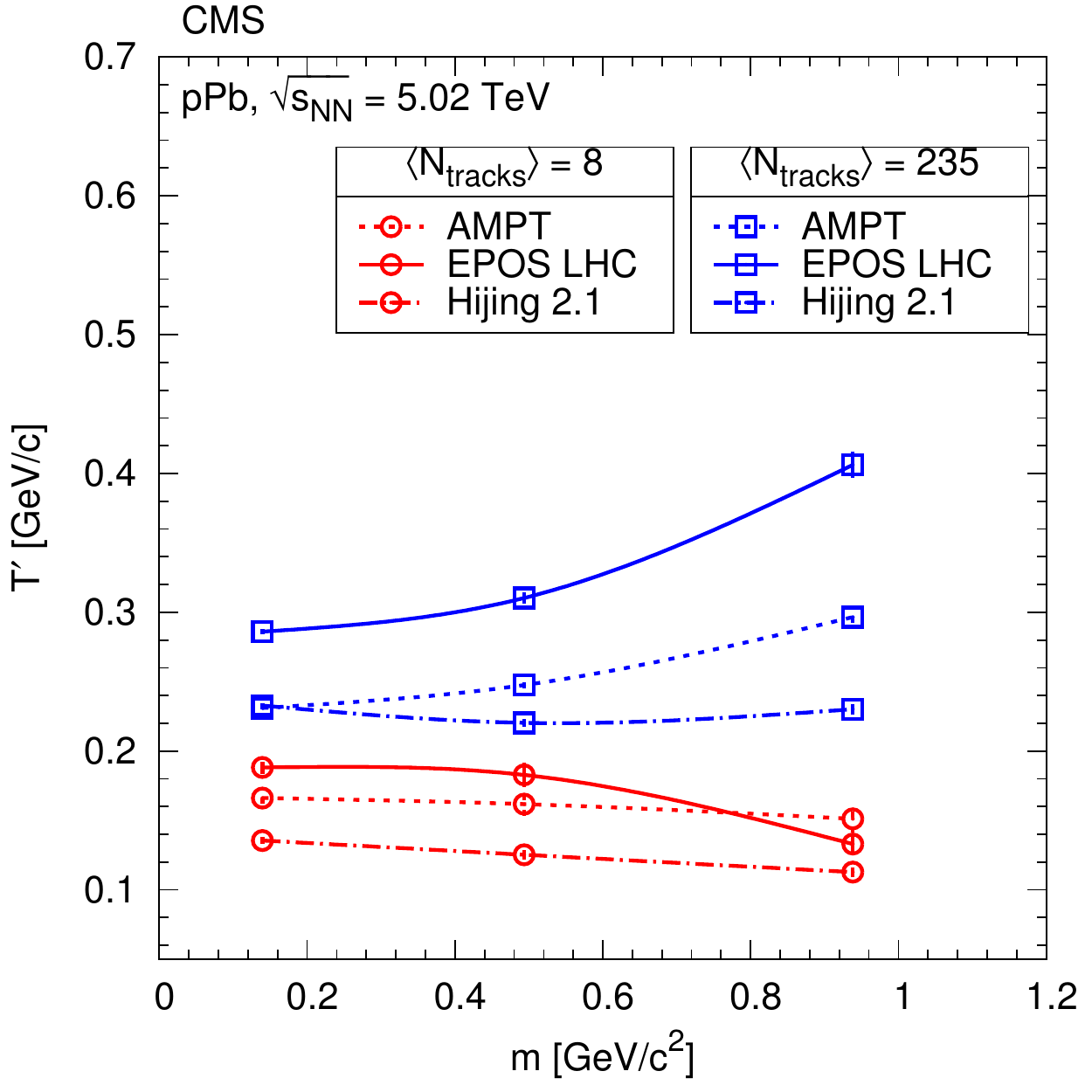}

 \caption{Inverse slope parameters $T'$ from fits of pion, kaon, and proton
spectra (both charges) with a form proportional to $\pt \exp(-\mt/T')$.
Results for a selection of multiplicity classes, with different $\langle
N_\text{tracks} \rangle$ as indicated, are plotted for pPb data (\cmsLeft) and
for MC event generators \AMPT, \EPOSLHC, and \HIJING (\cmsRight). The curves
are drawn to guide the eye. }

 \label{fig:invslope}

\end{figure}

For low track multiplicity ($N_\text{tracks} \lesssim 40$), pPb collisions
behave very similarly to pp collisions, while at higher multiplicities
($N_\text{tracks} \gtrsim 50$) the $\langle\pt\rangle$ is lower for pPb than in
pp.
The first observation can be explained since low-multiplicity events are
peripheral pPb collisions in which only a few proton-nucleon collisions are
present.
Events with more particles are indicative of collisions in which the projectile
proton strikes the thick disk of the lead nucleus.
Interestingly, the pPb curves (Fig.~\ref{fig:multiplicityDependence}, \cmsLeft
panel) can be reasonably approximated by taking the pp values and multiplying
their $N_\text{tracks}$ coordinate by a factor of 1.8, for all particle types.
In other words, a pPb collision with a given $N_\text{tracks}$ is similar to a
pp collision with $0.55 \times N_\text{tracks}$ for produced charged particles
in the $\abs{\eta} < 2.4$ range.
Both the highest-multiplicity pp and pPb interactions yield higher
$\langle\pt\rangle$ than seen in central PbPb collisions. While in the PbPb
case even the most central collisions possibly contain a mix of soft
(lower-$\langle \pt \rangle$) and hard (higher-$\langle \pt \rangle$)
nucleon-nucleon interactions, for pp or pPb collisions the most violent
interaction or sequence of interactions are selected.

\begin{figure}[htbp]

 \centering
  \includegraphics[width=0.49\textwidth]{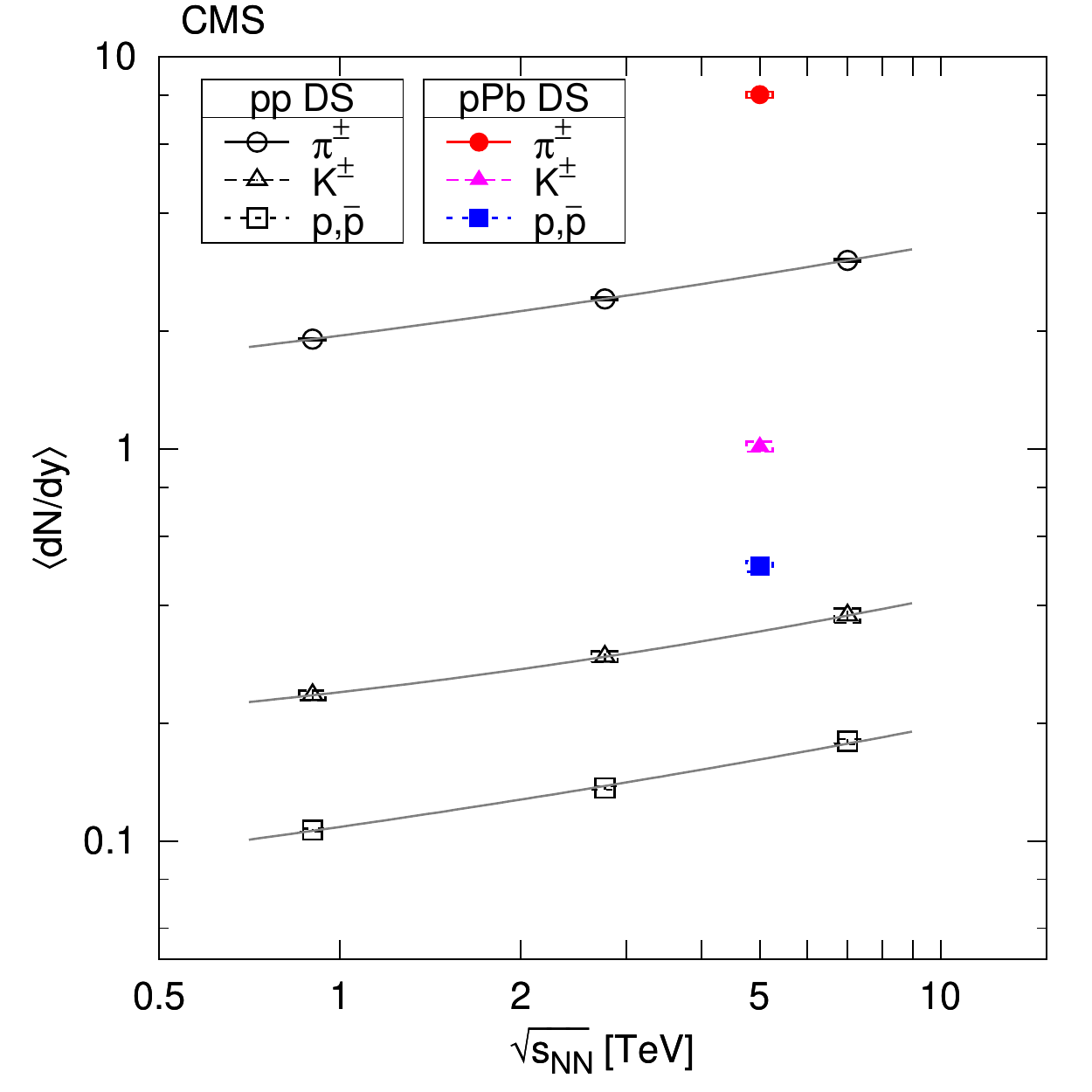}
  \includegraphics[width=0.49\textwidth]{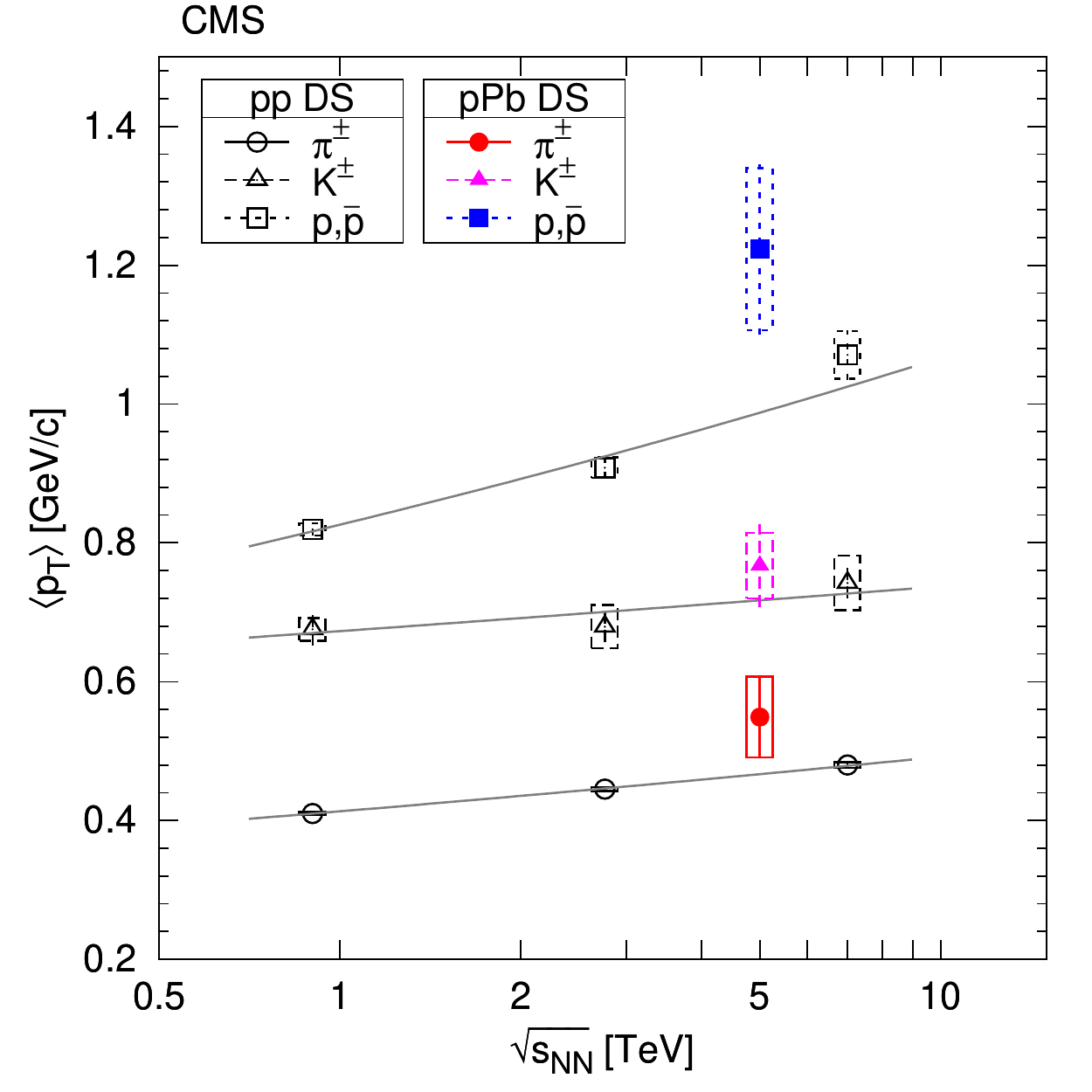}

 \caption{Average rapidity densities $\langle\rd N/\rd y\rangle$ (\cmsLeft) and average transverse
momenta $\langle \pt \rangle$ (\cmsRight) as a function of center-of-mass energy
for pp~\cite{identifiedSpectra} and pPb collisions, for charge-averaged pions,
kaons, and protons. Error bars indicate the uncorrelated combined
uncertainties, while boxes show the uncorrelated systematic uncertainties. The
curves show parabolic (for $\langle\rd N/\rd y\rangle$) or linear (for $\langle \pt \rangle$) interpolation on a log-log scale.
The pp and pPb data are for laboratory rapidity $\abs{y}<1$, which is the same
as the center-of-mass rapidity only for the pp data.
}

 \label{fig:energyDep}

\end{figure}

The transverse momentum spectra could also be successfully fitted
($\chi^2/\mathrm{ndf}$ in the range 0.7--1.8) with a functional form
proportional to $\pt \exp(-\mt/T')$, where $T'$ is called the inverse slope
parameter, motivated by the success of Boltzmann-type distributions in
nucleus-nucleus collisions~\cite{BraunMunzinger:2001ip}. In the case of pions,
the fitted range was restricted to $\mt > 0.4\GeVc$ in order to exclude the
region where resonance decays would significantly contribute to the measured
spectra.
The inverse slope parameter as a function of hadron mass is shown in
Fig.~\ref{fig:invslope}, for a selection of event classes, both for pPb data
and for MC event generators (\AMPT, \EPOSLHC, and \HIJING).
While the data display a linear dependence on mass with a slope that increases
with particle multiplicity, the models predict a flat or slowly rising behavior
versus mass and only limited changes with track multiplicity.
This is to be compared with pp results~\cite{identifiedSpectra}, where both
data and several tunes of the \PYTHIA~6~\cite{Sjostrand:2006za} and \PYTHIA~8
event generators show features very similar to those in pPb data.
A similar trend is also observed in nucleus-nucleus
collisions~\cite{Adler:2003cb,Abelev:2008ab}, which is attributed to the effect
of radial flow velocity boost~\cite{Schnedermann:1993ws}.

Average rapidity densities $\langle\rd N/\rd y\rangle$ and average transverse
momenta $\langle \pt \rangle$ of charge-averaged pions, kaons, and protons as a
function of center-of-mass energy are shown in Fig.~\ref{fig:energyDep} for pp
and pPb collisions, both corrected to the DS selection.
To allow comparison at the pPb energy, a parabolic (linear) interpolation of
the pp collision values at $\sqrt{s}= 0.9$, 2.76, and 7\TeV is shown for $\rd
N/\rd y$ $(\langle \pt \rangle)$.
The rapidity densities are generally about three times greater than in pp
interactions at the same energy, while the average transverse momentum
increases by about 20\%, 10\%, and 30\% for pions, kaons, and protons,
respectively.
The factor of three difference in the yields for pPb as compared to pp can be
compared with the estimated number of projectile collisions $N_\text{coll}/2 =
3.5 \pm 0.3$ or with the number of nucleons participating in the collision
$N_\text{part}/2 = 4.0 \pm 0.3$, based on the ratio of preliminary pPb and pp
cross-section measurements, that have proven to be good scaling variables in
proton-nucleus collisions at lower energies~\cite{PhysRevD.22.13}.

\section{Conclusions}

\label{sec:conclusions}

Measurements of identified charged hadron spectra produced in pPb collisions at
$\sqrt{s_{NN}} =5.02\TeV$ have been presented, normalized to events with
simultaneous hadronic activity at pseudorapidities $-5 < \eta < -3$ and $3 <
\eta < 5$. Charged pions, kaons, and protons were identified from the energy
deposited in the silicon tracker and other track information.
In the present analysis, the yield and spectra of identified hadrons for
laboratory rapidity $\abs{y} < 1$ have been studied as a function of the event
charged particle multiplicity in the range $\abs{\eta}<2.4$.
The $\pt$ spectra are well described by fits with the Tsallis-Pareto
parametrization. The ratios of the yields of oppositely charged particles are
close to one, as expected at mid-rapidity for collisions of this energy.
The average $\pt$ is found to increase with particle mass and the event
multiplicity.
These results are valid under the assumption that the particle yield
distributions follow the Tsallis-Pareto function in the unmeasured $\pt$
regions.

The results can be used to further constrain models of hadron production and
contribute to the understanding of basic non-perturbative dynamics in hadron
collisions.
The \EPOSLHC event generator reproduces several features of the measured
distributions, a significant improvement from the previous version, attributed
to a new viscous hydrodynamic treatment of the produced particles.
Other studied generators (\AMPT, \HIJING) predict steeper \pt distributions and
much smaller $\langle\pt\rangle$ than found in data, as well as substantial
deviations in the $\Pp/\Pgp$ ratios.

Combined with similar results from pp collisions, the track multiplicity
dependence of the average transverse momentum and particle ratios indicate that
particle production at LHC energies is strongly correlated with event particle
multiplicity in both pp and pPb interactions.
For low track multiplicity, pPb collisions appear similar to pp collisions.
At high multiplicities, the average $\pt$ of particles from pPb collisions with
a charged particle multiplicity of $N_\text{tracks}$ (in $\abs{\eta}<2.4$) is
similar to that for pp collisions with $0.55 \times N_\text{tracks}$.
Both the highest-multiplicity pp and pPb interactions yield higher
$\langle\pt\rangle$ than seen in central PbPb collisions.

\section*{Acknowledgments}

{\tolerance=800\hyphenation{Bundes-ministerium Forschungs-gemeinschaft Forschungs-zentren} We
congratulate our colleagues in the CERN accelerator departments for the
excellent performance of the LHC and thank the technical and administrative
staffs at CERN and at other CMS institutes for their contributions to the
success of the CMS effort. In addition, we gratefully acknowledge the computing
centres and personnel of the Worldwide LHC Computing Grid for delivering so
effectively the computing infrastructure essential to our analyses. Finally, we
acknowledge the enduring support for the construction and operation of the LHC
and the CMS detector provided by the following funding agencies: the Austrian
Federal Ministry of Science and Research and the Austrian Science Fund; the
Belgian Fonds de la Recherche Scientifique, and Fonds voor Wetenschappelijk
Onderzoek; the Brazilian Funding Agencies (CNPq, CAPES, FAPERJ, and FAPESP);
the Bulgarian Ministry of Education, Youth and Science; CERN; the Chinese
Academy of Sciences, Ministry of Science and Technology, and National Natural
Science Foundation of China; the Colombian Funding Agency (COLCIENCIAS); the
Croatian Ministry of Science, Education and Sport; the Research Promotion
Foundation, Cyprus; the Ministry of Education and Research, Recurrent financing
contract SF0690030s09 and European Regional Development Fund, Estonia; the
Academy of Finland, Finnish Ministry of Education and Culture, and Helsinki
Institute of Physics; the Institut National de Physique Nucl\'eaire et de
Physique des Particules~/~CNRS, and Commissariat \`a l'\'Energie Atomique et
aux \'Energies Alternatives~/~CEA, France; the Bundesministerium f\"ur Bildung
und Forschung, Deutsche Forschungsgemeinschaft, and Helmholtz-Gemeinschaft
Deutscher Forschungszentren, Germany; the General Secretariat for Research and
Technology, Greece; the National Scientific Research Foundation, and National
Office for Research and Technology, Hungary; the Department of Atomic Energy
and the Department of Science and Technology, India; the Institute for Studies
in Theoretical Physics and Mathematics, Iran; the Science Foundation, Ireland;
the Istituto Nazionale di Fisica Nucleare, Italy; the Korean Ministry of
Education, Science and Technology and the World Class University program of
NRF, Republic of Korea; the Lithuanian Academy of Sciences; the Mexican Funding
Agencies (CINVESTAV, CONACYT, SEP, and UASLP-FAI); the Ministry of Science and
Innovation, New Zealand; the Pakistan Atomic Energy Commission; the Ministry of
Science and Higher Education and the National Science Centre, Poland; the
Funda\c{c}\~ao para a Ci\^encia e a Tecnologia, Portugal; JINR (Armenia,
Belarus, Georgia, Ukraine, Uzbekistan); the Ministry of Education and Science
of the Russian Federation, the Federal Agency of Atomic Energy of the Russian
Federation, Russian Academy of Sciences, and the Russian Foundation for Basic
Research; the Ministry of Science and Technological Development of Serbia; the
Secretar\'{\i}a de Estado de Investigaci\'on, Desarrollo e Innovaci\'on and
Programa Consolider-Ingenio 2010, Spain; the Swiss Funding Agencies (ETH Board,
ETH Zurich, PSI, SNF, UniZH, Canton Zurich, and SER); the National Science
Council, Taipei; the Thailand Center of Excellence in Physics, the Institute
for the Promotion of Teaching Science and Technology of Thailand and the
National Science and Technology Development Agency of Thailand; the Scientific
and Technical Research Council of Turkey, and Turkish Atomic Energy Authority;
the Science and Technology Facilities Council, UK; the US Department of Energy,
and the US National Science Foundation.

Individuals have received support from the Marie-Curie programme and the
European Research Council and EPLANET (European Union); the Leventis
Foundation; the A. P. Sloan Foundation; the Alexander von Humboldt Foundation;
the Belgian Federal Science Policy Office; the Fonds pour la Formation \`a la
Recherche dans l'Industrie et dans l'Agriculture (FRIA-Belgium); the Agentschap
voor Innovatie door Wetenschap en Technologie (IWT-Belgium); the Ministry of
Education, Youth and Sports (MEYS) of Czech Republic; the Council of Science
and Industrial Research, India; the Compagnia di San Paolo (Torino); the HOMING
PLUS programme of Foundation for Polish Science, cofinanced by EU, Regional
Development Fund; and the Thalis and Aristeia programmes cofinanced by EU-ESF
and the Greek NSRF.
\par}

\bibliography{auto_generated}
\cleardoublepage \appendix\section{The CMS Collaboration \label{app:collab}}\begin{sloppypar}\hyphenpenalty=5000\widowpenalty=500\clubpenalty=5000\input{HIN-12-016-authorlist.tex}\end{sloppypar}
\end{document}

%% file: HIN-12-016-authorlist.tex
\textbf{Yerevan Physics Institute,  Yerevan,  Armenia}\\*[0pt]
S.~Chatrchyan, V.~Khachatryan, A.M.~Sirunyan, A.~Tumasyan
\vskip\cmsinstskip
\textbf{Institut f\"{u}r Hochenergiephysik der OeAW,  Wien,  Austria}\\*[0pt]
W.~Adam, T.~Bergauer, M.~Dragicevic, J.~Er\"{o}, C.~Fabjan\cmsAuthorMark{1}, M.~Friedl, R.~Fr\"{u}hwirth\cmsAuthorMark{1}, V.M.~Ghete, N.~H\"{o}rmann, J.~Hrubec, M.~Jeitler\cmsAuthorMark{1}, W.~Kiesenhofer, V.~Kn\"{u}nz, M.~Krammer\cmsAuthorMark{1}, I.~Kr\"{a}tschmer, D.~Liko, I.~Mikulec, D.~Rabady\cmsAuthorMark{2}, B.~Rahbaran, C.~Rohringer, H.~Rohringer, R.~Sch\"{o}fbeck, J.~Strauss, A.~Taurok, W.~Treberer-Treberspurg, W.~Waltenberger, C.-E.~Wulz\cmsAuthorMark{1}
% ...
\vskip\cmsinstskip
% ...
1:~~Also at Vienna University of Technology, Vienna, Austria\\
2:~~Also at CERN, European Organization for Nuclear Research, Geneva, Switzerland\\
% ...
